\documentclass{aastex7}
\received{May 26, 2025}
\revised{July 22; September 9; September 18, 2025}
\accepted{September 24, 2025}

\submitjournal{AJ}

\begin{document}


\title{Characterizing Temperatures of Flares on the M Dwarf Wolf~359 from Simultaneous Multiband Optical Observations}

\correspondingauthor{Chia--Lung Lin}
\author[0000-0001-5989-7594]{Chia--Lung Lin}
\affiliation{Graduate Institute of Astronomy, National Central University, Taoyuan 32001, Taiwan}
\affiliation{Steward Observatory, The University of Arizona, Tucson, AZ 85721, USA}
\email{m1059006@gm.astro.ncu.edu.tw, chialunglin@arizona.edu}

\author[0000-0002-9679-5279]{Li--Ching Huang}
\affiliation{Shanghai Astronomical Observatory, Chinese Academy of Sciences, Shanghai, China}
\email{lchuang@ntnu.edu.tw}

\author{Wei--Jie Hou}
\affiliation{Graduate Institute of Astronomy, National Central University, Taoyuan 32001, Taiwan}
\email{weij@astro.ncu.edu.tw}

\author{Hsiang--Yao Hsiao}
\affiliation{Graduate Institute of Astronomy, National Central University, Taoyuan 32001, Taiwan}
\email{durian@gm.astro.ncu.edu.tw}

\author[0000-0002-3140-5014]{Wing--Huen Ip}
\affiliation{Graduate Institute of Astronomy, National Central University, Taoyuan 32001, Taiwan}
\affiliation{Graduate Institute of Space Science, National Central University, Taoyuan 32001, Taiwan}
\email{wingip@astro.ncu.edu.tw}




\begin{abstract}

{We present a flare temperature study of the highly active M~dwarf Wolf~359 using simultaneous multiband ($u$, $g$, $r$, $i$, and $z$) photometric observations from the Lulin 1-m and 41-cm telescopes.  
Twelve flares were detected over five nights, with significant brightness increases in the $u$, $g$, and $r$~bands; only three were seen in $i$, and none in $z$.  
From broadband SED fitting and $g$/$r$ color ratio, we derive an average flare temperature of $5500 \pm 1600$~K, significantly cooler than the canonical 10000~K.  
{We obtained a power-law relation between FWHM flare temperature and energy in the solar-class flare regime and extrapolated it to higher energies, superflare regime. 
This power-law is consistent with the trends reported for M-dwarf superflares in previous studies, suggesting a common temperature–energy scaling across several orders of magnitude. 
However, the scatter in the superflare regime increases, indicating that such energetic events may involve more complex physical mechanisms and limiting the applicability of simple blackbody models at the high energy flares.}
Using our {FWHM flare temperature--TRIPOL~$g$ energy relation} and the reported flare energy frequency distribution of Wolf~359, we evaluated the potential flare contribution to photosynthetically active radiation (PAR) in the habitable zone.  
{We find that typical solar-class giant flares ($E_{\mathrm{fl,bol}} \sim 9\times10^{31}$~erg, $T_{\mathrm{fl,fwhm}} \sim 6800$~K) are {not frequent enough} to sustain Earth-like net primary productivity.  
Even under the extreme superflare condition ($\sim$$10^{36}$~erg, $\sim$16500~K), flare activity remains far from meeting the PAR threshold.}
}

\end{abstract}


\section{Introduction} \label{sec:intro}

Stellar flares are sudden and powerful energy releases caused by magnetic reconnection in a star's atmosphere \citep[e.g.,][]{1963ApJS....8..177P}. 
This process leads to rapid heating of the surrounding plasma and accelerates charged particles along magnetic loops, producing emission across the electromagnetic spectrum, from high-energy X-rays to low-frequency radio waves. 
Flares are also often accompanied by particle outburst, such as coronal mass ejections (CMEs) and stellar proton events (SPEs), which further contribute to the energetic impact of these events {\citep[][]{2005ApJ...621..398O, 2015A&A...581A..28P, 2019ApJ...871..167K, 2023MNRAS.519.3564J, 2024ApJ...961...23N,2025ApJ...982...43B}.}
{Based on their total released energy, flares can be categorized into {solar-class} flares with energies comparable to typical solar flares in the range of $10^{29}$ to $10^{32}$~erg, and superflares with energies more than {$10^{33}$~erg} \citep[e.g.,][]{2012Natur.485..478M}.}

Space-based photometric surveys such as the Kepler/K2 missions \citep[][]{2010Sci...327..977B,2010ApJ...713L..79K, 2014PASP..126..398H} and the Transiting Exoplanet Survey Satellite (TESS) \citep{2015JATIS...1a4003R} have revolutionized our capacity to monitor transient flare events. 
These datasets have enabled the detection of energetic superflares across a broad range of stellar types, from A-type to {L-type dwarfs \citep{2012Natur.485..478M, 2012MNRAS.423.3420B, 2013ApJS..209....5S, 2014ApJ...797..121H, 2015ApJ...798...92W, 2017ApJ...845...33G, 2018ApJ...867...78C, 2019ApJS..241...29Y, 2019ApJ...873...97L, 2018ApJ...858...55P, 2020MNRAS.494.5751P, 2020AJ....159...60G, 2020ApJ...905..107M, 2023A+A...669A..15Y, 2024MNRAS.527.8290P, 2024AJ....168..234L}.}
These findings affirm that magnetic reconnection-driven flares are widespread among main-sequence stars, particularly those with substantial outer convective zones. 
Among them, cooler dwarfs (spectral types M and K) demonstrate significantly elevated flare occurrence rates \citep[e.g.,][]{2019ApJ...873...97L, 2023A+A...669A..15Y, 2024AJ....168..234L}, a trend that is thought to reflect their underlying magnetic dynamo efficiency.

Furthermore, the enhanced ultraviolet radiation and associated CME and SPE activity produced by superflares may strongly influence the atmospheres of nearby exoplanets by triggering photochemical reactions that modify their composition \citep{2010AsBio..10..751S, 2021NatAs...5..298C, 2025arXiv250503723C}.
Flares also pose significant challenges for exoplanet characterization. The photometric and spectroscopic variability induced by flare events can interfere with the detection of planetary transits and introduce systematic errors in radial velocity measurements. 
In addition, flare-induced emission can contaminate near-infrared transmission spectra, complicating efforts to accurately retrieve the atmospheric composition of exoplanets \citep{2018ApJ...853..122R, 2018AJ....156..178Z, 2023AJ....165..149D, 2023ApJ...959...64H}.


Flare temperature is a key parameter for determining the bolometric energy of a flare \citep[e.g.,][]{2013ApJS..209....5S}.
It is also critical for estimating the associated ultraviolet (UV) flux and excess emission {\citep[][]{2023MNRAS.519.3564J, 2024ApJ...971...24P}}, for evaluating potential impacts on the habitability of Earth-like planets in the habitable zone \citep[e.g.,][]{2019MNRAS.485.5924L, 2020AJ....159...60G}, and for potentially modeling stellar contamination in NIR exoplanet transmission spectra \citep[e.g.,][]{2023ApJ...959...64H}.

For decades, {flare energies} have typically been estimated assuming blackbody continuum emission temperatures of 9000--10000~K, derived from multi-band photometric and spectroscopic observations of white-light flares on a handful of well-studied M dwarfs.  
These temperatures were obtained by fitting broadband spectral energy distributions (SEDs) or optical spectra in the wavelength range of 4000–4800~\AA\ {\citep[e.g.,][]{2003ApJ...597..535H}.}
However, recent studies have called this constant temperature assumption into question.  
For example, \cite{2023MNRAS.519.3564J} has shown that the widely adopted 9000~K blackbody model significantly underestimates M dwarf flare emission in the ultraviolet, by up to a factor of 6.5 in the near-ultraviolet (NUV) and up to a factor of 10 in the far-ultraviolet (FUV).
{Similarly, \citet{2024ApJ...971...24P} found that the 9000~K blackbody model underestimates the NUV flare energies for events exceeding $10^{30}$~erg by a factor of approximately 2.5.
These studies both highlights the need for revised flare temperature models when estimating the UV radiation environments of low-mass stars.}

\citet{2020ApJ...902..115H} analyzed simultaneous two-band observations of flares from several low-mass stars using TESS and Evryscope \citep[][]{2019PASP..131g5001R}, and found that flare temperature evolves throughout the flaring phase.  
{They also reported a positive correlation between flare energy and the FWHM flare temperature—defined as the average temperature within the full width at half maximum of the flare light curve—for low-mass stars from mid-K to M type, primarily in the superflare regime, with temperatures ranging from 6000 K 40,000 K.}
{However, when considering only the M~dwarfs with masses $< 0.52M_{\odot}$ from their data, the positive correlation between superflare temperature and energy becomes less significant, with an average flare temperature of $T_{\mathrm{fl}} = 9800^{+3300}_{-2400}$~K.}
Similarly, {\cite{2024MNRAS.529.4354J}} reported a peak flare temperature of  $7100^{+150}_{-130}$~K for a multiband-observed superflares on the M-dwarf binary CR~Dra. 
Beyond late-type low-mass stars, \citet{2022AJ....164..223R} analyzed over 200 stellar flares in F- to K-type stars observed by the \textit{CoRoT} satellite and derived their blackbody temperatures using color photometry, finding a wide temperature range (3000–300000~K) with a distribution peaking near 6400~K at spectral type of G6.

Consistent with these, \citet{Maas+2022} conducted multiband observations of flares on TRAPPIST-1 and measured the peak flare temperatures of $T_{\mathrm{fl}} = 8290^{+660}_{-550}$~K and $T_{\mathrm{fl}} = 13620^{+1520}_{-1220}$~K for the flares with lower, solar-class energy of $\sim 10^{31}$~erg.
More recently, JWST near-infrared spectroscopic observations of TRAPPIST-1 revealed that the continuum emission of smaller flares ($\lesssim 10^{31}$~erg) is well described by blackbody models with even cooler peak temperatures, {below 5300~K} \citep[][]{2023ApJ...959...64H}, significantly lower than the canonical 9000–10000~K.
Together, these studies extend the flare temperature–energy correlation into the low-energy solar-class flare regime, where the positive trend begins to emerge more clearly.

Nevertheless, the number of flare temperature measurements in the low-energy regime remains limited, and additional observations are needed to robustly characterize the correlation. 
To improve our understanding of the flare temperature–energy correlation of M~dwarfs' flares, especially at lower energies, we carried out a multiband photometric monitoring campaign of Wolf~359, an extremely flare-active late-M dwarf, using the Lulin 1-m and 41-cm telescopes in Taiwan.  
This setup enabled us to capture flares simultaneously in up to five optical bands ($u$, $g$, $r$, $i$, and $z$), allowing us to derive temperature-sensitive color indices and track the evolution of flare continuum properties with high cadence.

The structure of this paper is organized as follows:  
Section~\ref{sec:observations} introduces our target, Wolf~359, and describes the observational setup, instruments used, and the observing strategy.  
In Section~\ref{sec:Data processing}, we carry out the data reduction and detail the differential photometry techniques applied to calibrate the extracted light curves of Wolf~359.  
Section~\ref{sec:flare detection and analysis} outlines the methods used for flare detection, and for estimating flare energies and temperatures.  
The results of our analysis are presented in Section~\ref{sec:results}.  
{In Section~\ref{sec:discussion}, we examine the temperature properties of the flares detected in this study and discuss the implications for exoplanetary habitability based on our derived flare temperature-energy correlation and the flare frequency distribution reported in literature.}
Finally, Section~\ref{sec:conclusions} summarizes our findings and outlines potential directions for future work.

\section{Observation}
\label{sec:observations}
Wolf~359 is one of the nearest late-type active M dwarfs, located at a distance of 2.42~pc \citep{2016AJ....152...24W}, and is classified as spectral type M5.5 \citep{1995AJ....110.1838R} to M6 \citep{1991ApJS...77..417K}.  
{The stellar radius of Wolf~359 is $R_* = 0.16~R_{\odot}$ \citep[][]{stellar_radius}.}
Its rotation period of $2.72 \pm 0.04$~days \citep{2018RNAAS...2....1G} indicates {a young gyrochronological age between 0.1 and 1.5~Gyr \citep[][]{2023AJ....166..260B}.}
The star has an estimated effective temperature between 2700--2900~K \citep{effective_temperature_1, effective_temperature_2} and magnitudes of $g = 14.265$, $r = 12.762$, and $i = 10.320$ \citep[][]{2012yCat.1322....0Z}, making it relatively bright in the red optical and near-infrared bands.  
Like other late-type M dwarfs, Wolf~359 possesses a strong surface magnetic field of approximately 2.4~kG \citep{2007ApJ...656.1121R}, which drives frequent and energetic flaring activity.  
The {white light} flare occurrence rate has been observed to reach up to 11 giant flares per day \citep[][]{2021AJ....162...11L}, making Wolf~359 an ideal target for time-limited flare monitoring campaigns.
{A summary of its stellar parameters is provided in Table~\ref{tab:wolf359_summary}.}


We conducted our observation campaign of Wolf~359 over five nights during the dark moon phase, from February 17 to 22, 2023, at Lulin Observatory in Taiwan. 
The observation scheduled for February 19th was canceled due to poor weather conditions.
The campaign utilized two instruments: the three-color Triple Range Imager and POLarimeter \citep[TRIPOL,][]{2019RAA....19..136S}, installed on the Lulin 1-m telescope, and the photometer on the Lulin 41-cm telescope.

TRIPOL is capable of simultaneous photometric observations in three optical bands: {$g^{}$, $r^{}$, and $i^{}$ with the effective wavelengths ($\lambda_{eff}$) of $\sim$~4750~$\AA$, 6200~$\AA$, and 7700~$\AA$, respectively.}
It was specifically designed for 1-m class ground-based telescopes to perform multicolor follow-up and time-series observations of color-dependent phenomena, such as the temperature variability of stellar flares.  
It is equipped with three $512 \times 512$ CCDs with a pixel scale of 0.47 arcseconds per pixel, resulting in a field of view of $4 \times 4$ arcmin$^2$.

The Lulin 41-cm telescope is equipped with a $1024 \times 1024$ CCD and provides a larger field of view of $27 \times 27$ arcmin$^2$.  
It is also equipped with a fast filter wheel, allowing it to switch between two different filters between exposures.  
{We originally planned to employ the $u$ ($\lambda_{eff}$=3540~$\AA$) and $z$~bands ($\lambda_{eff}$=10084~$\AA$) \footnote{See \url{https://www.lulin.ncu.edu.tw/instrument/SLT/}.}
 on the Lulin 41-cm telescope to achieve five~band simultaneous flare observations of Wolf~359.}
Unfortunately, this was only possible on the first night of observation. The $u$~band filter malfunctioned on the second night and could not be repaired in a short time, so only the $z$~band was used for the remaining observation runs.


Each night, we collected bias frames, dark frames, and dome flat frames at the beginning and end of the run for subsequent data reduction.  
During the scientific observations, we continuously monitored Wolf~359 for around {9~hours each night to maximize the chance of capturing as many flare events as possible.}

The observation cadence, including the read-out time, with TRIPOL was set to approximately 35~seconds {(30~seconds exposure $+$ 5~seconds readout)} for the runs on February 17 and 18.  
Starting from February 20, due to the very dark moon phase, we reduced the cadence to 25~seconds {(20~seconds exposure $+$ 5~seconds readout)} and maintained it for the remainder of the observations.

{On the first night, February 17, observations in the $u$ and $z$ bands using the Lulin 41-cm telescope involved alternating between a 300~seconds exposure in the $u$~band and a 5~seconds exposure in the $z$~band.
Each cycle also included 30~seconds of readout time and approximately 29~seconds for filter wheel switching, resulting in a total cadence of approximately 364~seconds.}
From February 18 onward, due to the failure of the $u$~band filter, the Lulin 41-cm telescope used only the $z$~band for observations, with a cadence of about 20~seconds ({5~seconds exposure $+$ 15~seconds readout}).
The summary of this observation campaign are provided in Table~\ref{tab:obs_summary}.

\section{Data reduction, light curve extraction and calibration}
\label{sec:Data processing}
We reduced the dataset using a customized Python pipeline following the standard IRAF photometric reduction procedures, including bias subtraction, dark subtraction, and flat-fielding correction.
The integrated fluxes of point sources in the images were extracted using \texttt{SExtractor} \citep[][]{1996A&AS..117..393B} through aperture photometry.

\begin{figure}
    \centering
    \epsscale{1}
    \plotone{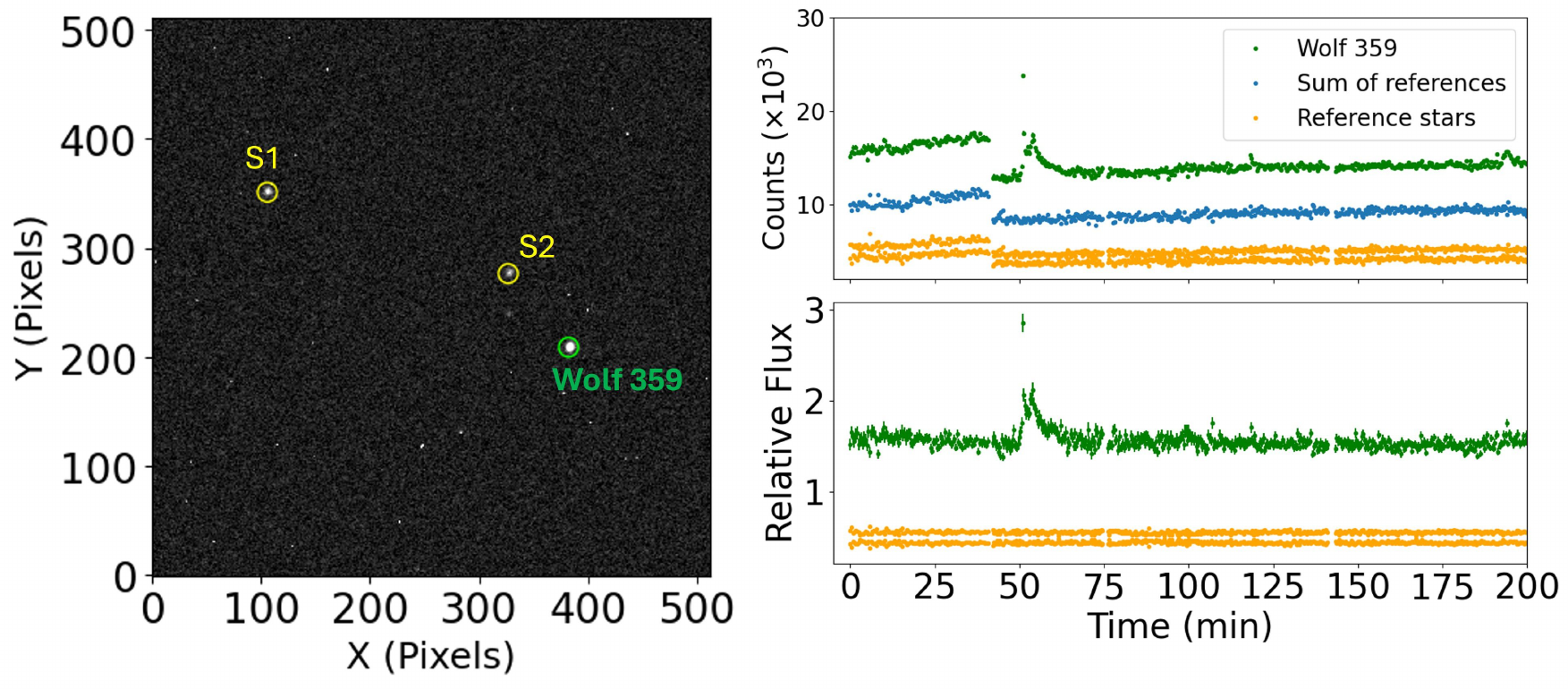}
    \caption{Left: One of the TRIPOL $g$~band images of Wolf~359 taken on February 22, 2023, with the target and two selected differential standard stars (S1 and S2) labeled.  
    Right, top panel: Raw light curves of Wolf~359 (green), the sum of the two reference stars (blue), and the individual reference stars (orange), in units of instrumental counts.  
    Right, bottom panel: Calibrated relative flux light curves from differential photometry. A flare event is clearly visible in Wolf~359 around 60 minutes after the start of the observation run.  
    The two reference stars exhibit minimal variability, validating their use as stable differential standards. 
    The same processing was applied to the data obtained in all other bands and on all other observation nights.}
    \label{fig:differential_photometry}
\end{figure}

For each optical band of data, we extracted the raw light curve of Wolf~359 as well as the light curves of nearby field stars to serve as differential standards.  
Due to the small field of view of TRIPOL, only two field stars could be selected as suitable differential standards.  
We calibrated the atmospheric extinction and variability in the raw light curve of Wolf~359 by performing differential photometry using these two reference stars.
Figure~\ref{fig:differential_photometry} shows an image of Wolf~359 and the two standard stars taken by TRIPOL in the $g$~band on February 22, 2023, along with the calibrated light curves of Wolf~359 obtained via differential photometry as an example.  
A flare was captured at approximately 60~minutes after the start of the observation run.
We confirmed that these two standard stars exhibit significantly less variability than Wolf~359, making them reliable differential standards that do not introduce contamination into the calibrated light curve or the detected flare events.

For the Lulin 41-cm telescope data with a larger field of view, we selected six field stars, including the two used for the TRIPOL data, as differential references to calibrate the raw light curves of Wolf~359 via differential photometry in the $u$ and $z$~bands.

\section{Flare detection, analysis, and temperature estimation.}
\label{sec:flare detection and analysis}
\subsection{Flare detection}
For light curves observed in all bands in each night, we normalized the light curves, $\Delta F(t)$, by dividing by the their median fluxes, $\widetilde{F}$,
\begin{equation}
  \frac{\Delta F(t)}{\widetilde{F}} = \frac{F(t)-\widetilde{F}}{\widetilde{F}}.
  \label{eq:eq_amplitude}
\end{equation}
We first searched for flares in the $g$~band light curves and then identified their counterparts in the light curves of other bands.  
To isolate the flare signals, we visually selected data points from time intervals without evident flares and generated a flare-free baseline using quadratic interpolation.  
This interpolated baseline was then subtracted from the $g$~band light curve to produce a residual light curve from which flare events could be identified.


We adopted the identification criteria from \citet[][]{2024AJ....168..234L}, a modified version of the method proposed by \citet{2015ApJ...814...35C}, to detect flare events in our data.  
{Specifically, we required: (1) a positive normalized flux deviation, (2) a signal-to-noise ratio of at least 1 relative to the standard deviation of the light curve, and (3) a minimum of three consecutive data points meeting these conditions, in order to avoid false positives due to noise fluctuations.}
Using this approach, we identified twelve flares in Wolf~359 over the five nights of $g$~band observations. 
We then applied the same method, along with visual inspection, to identify the $g$~band flare counterparts in the light curves of the other bands (Figure~\ref{fig:Feb17_flares}).  

Most flares exhibited significant amplitudes in the $u$, $g$, and $r$ bands.  
On the first night, February 17, the three flares detected in the $g$~band were also clearly seen in the $u$ and $r$ bands, and marginally in the $i$~band.  
For the rest of nights, all $g$~band flares had identifiable counterparts in the $r$~band, while only three were detectable in the $i$~band.  
None of the flares had detectable counterparts in the $z$~band.

The basic parameters of each flare, peak amplitude and duration, were also determined at this stage. 
The flare peak amplitude, representing the brightness increase in excess of the nominal stellar luminosity, was defined as the maximum value in the normalized light curve during the flare.  
The flare duration was measured as the time interval between the first and last data points with normalized flux exceeding the 1--$\sigma$ threshold.
{We adopted this relatively low threshold to mitigate the underestimation of flare duration that arises in light curves with limited photometric precision.
The photometric precision (1--$\sigma$ noise level) of our TRIPOL light curve data is about 0.01-0.02.
Higher thresholds (e.g., 2- or 3--$\sigma$) would exclude low-amplitude segments of the flare, particularly in the rise and decay phases, leading to systematically shorter measured durations.}

\subsection{Flare energy}
\label{subsec:flare_energy}
The energy of each flare can be characterized by its equivalent duration (ED), defined as the time required for the star in quiescence to emit the same amount of energy as that released by the flare. The equivalent duration is calculated following \citet{1972Ap&SS..19...75G}:
\begin{equation}
  ED = \int \frac{\Delta F(t)}{\widetilde F} dt . 
  \label{eq:ED}
\end{equation}
It is the integrated area under the normalized flare light curve, expressed in units of seconds.
The energy of a flare in a given band can be estimated using the following equation:
\begin{equation}
E_{f, \mathrm{band}} = ED \times L_{*, \mathrm{band}},
\label{eq:flare_energy}
\end{equation}
where $E_{f, \mathrm{band}}$ is the flare energy in the specified band, and $L_{*, \mathrm{band}}$ is the star's quiescent luminosity within the transmission of that band.



Wolf~359 is a late-M star, its quiescent spectrum exhibits strong TiO absorption features in the optical bands. 
Therefore, assuming an ideal blackbody with a temperature of 2800~K would significantly overestimate the star's optical luminosity in quiescence.  
Moreover, since Wolf~359 is an extremely active star, its brightness can be highly variable.

To ensure the accuracy of our measurements, we derived the quiescent luminosities of Wolf~359 in the $g$, $r$, and $i$ bands from observed spectra of the star.  
The spectra were obtained using the Lulin 1-m telescope with a low-resolution ($R \sim 1000$) optical spectrograph, Shelyak LISA {\citep[][]{thizy2010spectrographs}}, covering the wavelength range 4000--8400~\AA, on April 20, 2021.  
{A total of 20 spectra of Wolf~359 were obtained that night, each with an exposure time of 600~seconds. The average readout time per spectrum was approximately 5~seconds, resulting in a cadence of about 605~seconds. Simultaneous $u$~band photometric observations were conducted using the Lulin 41-cm telescope, with an exposure time of 180~seconds per frame and a readout time of 10~seconds, yielding a cadence of approximately 190~seconds (see Table~\ref{tab:obs_summary}).}

The spectroscopic standard star Feige~66 was used for absolute flux calibration.  
The spectral data were reduced and calibrated using standard IRAF procedures, including instrumental, wavelength, and flux calibration.
Figure~\ref{fig:quiescent_spectrum_wolf359} shows the $u$~band light curve of the observation, with dashed lines marking the midpoint of each spectroscopic exposure.  
Based on this light curve, we confirmed that spectra No.~7 through No.~14 were very likely not taken during a flare.  
We averaged these spectra to construct a representative quiescent spectrum (also shown in Figure~\ref{fig:quiescent_spectrum_wolf359}), which was then used to measure the quiescent luminosities in the optical bands.



From the observed quiescent spectrum of Wolf~359, we derived the average flux densities ($F_{\lambda}$) at the central wavelengths of the $g$, $r$, and $i$ bands to be $9.08\times10^{-15}~\mathrm{erg\,s^{-1}\,cm^{-2}\,\AA^{-1}}$, $2.26\times10^{-14}~\mathrm{erg\,s^{-1}\,cm^{-2}\,\AA^{-1}}$, and $1.76\times10^{-13}~\mathrm{erg\,s^{-1}\,cm^{-2}\,\AA^{-1}}$, respectively.  
These values yield bandpass-integrated fluxes ($F_{*}$), based on the FWHM of each filter's transmission curve, of $1.26\times10^{-11}~\mathrm{erg\,s^{-1}\,cm^{-2}}$ in the $g$ band, $2.65\times10^{-11}~\mathrm{erg\,s^{-1}\,cm^{-2}}$ in the $r$ band, and $2.43\times10^{-10}~\mathrm{erg\,s^{-1}\,cm^{-2}}$ in the $i$ band.
The corresponding luminosities ($L_{*}$) were then calculated using $L = 4\pi d^{2} F_{*}$, where $d$ is the distance to Wolf~359 in centimeters.  
As a result, the quiescent luminosities of Wolf~359 in the optical bands are $L_{*,g} = 3.37\times10^{27}~\mathrm{erg\,s^{-1}}$, $L_{*,r} = 1.03\times10^{28}~\mathrm{erg\,s^{-1}}$, and $L_{*,i} = 7.16\times10^{28}~\mathrm{erg\,s^{-1}}$.


For the $u$~band, we estimated the quiescent luminosity using the $u-g = 3.05$ color index reported by \citet{2007AJ....134.2398C} for M5 dwarf stars.  
Given that the $g$~band magnitude of Wolf~359 from our observation was approximately 14.30, its $u$~band magnitude was inferred to be 17.35.  
This magnitude was then converted to a flux density of $1\times10^{-15}~\mathrm{erg\,s^{-1}\,cm^{-2}\,\AA^{-1}}$, corresponding to a bandpass-integrated flux of $6\times10^{-13}~\mathrm{erg\,s^{-1}\,cm^{-2}}$ and a quiescent luminosity of $L_{*,u} = 4.25\times10^{26}~\mathrm{erg\,s^{-1}}$ in the $u$~band.

Finally, we estimated the flare energies in the different observed bands using the equivalent durations and the quiescent luminosities derived above by following Eq.~\ref{eq:flare_energy}.
{Table~\ref{tab:flare_properties} presents the complete set of energy measurements together with other flare parameters, such as duration and peak amplitude.}

\subsection{Flare temperature}
With the simultaneous multiband observations, the flare blackbody temperature can be inferred.  
We explored two methods for estimating the flare temperature:  
(1) fitting the flare-only spectral energy distribution (SED), and  
(2) using the color ratio between flux densities in two different bands.

\subsubsection{Flare-only SED method}
\label{subsubsec:flare-only_sed_method}

Balmer continuum emission is commonly observed in M dwarf flares, with varying contributions at wavelengths shorter than the Balmer jump ($\lambda < 3600$~\AA), where the $u$~band is most sensitive \citep[][]{2006ApJ...644..484A, 2010ApJ...714L..98K}.  
Therefore, we excluded the $u$~band data from the SED fitting method used to estimate flare temperatures.
In addition, the $z$~band data were excluded because of the lack of detectable flare amplitudes in this bandpass during our observations.  
We performed blackbody temperature estimates using the data in the $g$, $r$, and $i$ bands.

The flare-only flux density of the band, $F_{fl, \lambda, band}$, was calculated by multiplying the flare amplitudes with Wolf~359's quiescent flux density at the central wavelength of the corresponding bandpass, as derived in Section~\ref{subsec:flare_energy}:
\begin{equation}
F_{fl, \lambda, band} =  (\frac{\Delta F}{\widetilde{F}})_{band} \times F_{*, \lambda, band}.
\label{eq:flare_flux_density}
\end{equation}
We constrained the blackbody temperature of the flares by fitting the resulting flare-only SED with the following equation \citep[][]{2003ApJ...597..535H}
\begin{equation}
F_{fl,\lambda} = X_{\mathrm{BB}} \frac{R_*^2}{d^2} \pi B_{\lambda}(T_{fl}),
\label{eq:flare_temperature}
\end{equation}
where $B_{\lambda}$ is the Planck function of flare temperature $T_{fl}$ in Kelvin, and $X_{\mathrm{BB}}$ is the filling factor of the flare-heated region on the stellar visible surface.  

Figure~\ref{fig:sed_fitting_examples} shows the {instantaneous} peak flare-only SEDs and the best-fit blackbody models for the flares detectable in all $g$, $r$, and $i$ bands: {F5-0220 (Flare No.~1 on February 20), F10-0221 (Flare No.~4 on February 21), and F12-0222 (Flare No.~1 on February 22).
The corresponding peak temperatures are $T_{fl,\mathrm{SED}} = 5735 \pm 550$~K (F5-0220), $T_{fl,\mathrm{SED}} = 4549 \pm 370$~K (F10-0221), and $T_{fl,\mathrm{SED}} = 5469 \pm 450$~K.}  
These values are not only significantly cooler than the commonly adopted empirical temperature of 10000~K, but also fall below the average M dwarf flare temperature range reported by \citet{2021AJ....162...11L}, based on the dataset from \citet{2020ApJ...902..115H}.


\subsubsection{Color ratio method}

The flare {blackbody} temperature can be also derived from the the ratio of flux densities between these two bands using the equation expressed as
\begin{equation}
\frac{F_{fl,\lambda,\mathrm{band1}}}{F_{fl,\lambda,\mathrm{band2}}} = \frac{B_{\lambda,\mathrm{band1}}(T_{fl})}{B_{\lambda,\mathrm{band2}}(T_{fl})},
\label{eq:color_ratio}
\end{equation}
where $F_{fl,\lambda,\mathrm{band1}}$ and $F_{fl,\lambda,\mathrm{band2}}$ are the flare-only flux densities at the central wavelengths of the two bandpasses, computed using Equation~\ref{eq:flare_flux_density}, with $\mathrm{band1}$ being the bluer band.  
$B_{\lambda,\mathrm{band1}}(T_{fl})$ and $B_{\lambda,\mathrm{band2}}(T_{fl})$ represent the spectral radiance at the respective central wavelengths, given by the Planck function at flare temperature $T_{fl}$.
{We estimate the flare temperatures using this method for the same flares discussed in Section~\ref{subsubsec:flare-only_sed_method}. 
We summarize those temperatures in Table~\ref{tab:flare_temperature_comparison}.}
The flare temperatures derived from the $g$/$r$ ratio at their instantaneous peak brightness are generally consistent with those estimated from SED fitting, although with larger uncertainties.  
In contrast, the temperatures estimated from the $r$/$i$ and $g$/$i$ ratios are a bit more divergent from those obtained by the $g$ /$r$ ratio and SED fitting.
{One possible explanation is that M dwarfs are intrinsically brighter in the $i$~band than in the $g$ or $r$ bands, the relative flare amplitude in the $i$~band is smaller and more strongly affected by the higher baseline stellar flux and associated noise.   
As shown in Figure~\ref{fig:sed_fitting_examples}, the $i$~band fluxes tend to exhibit larger scatter due to photometric uncertainties and the elevated background, which reduces the reliability of temperatures derived from two-band flux ratios involving the $i$ band.   
Consequently, the $r$/$i$ and $g$/$i$ flux ratios are more susceptible to measurement errors, resulting in less accurate temperature estimates compared to those obtained from the $g$/$r$ ratio or triple-band SED.}

Given these results, and considering that most of the flares detected in our TRIPOL observations were only clearly detectable in the $g$ and $r$ bands, we adopted the $g$/$r$ ratio method to estimate blackbody temperatures for the remaining flares analyzed in this study.  

\section{Results}
\label{sec:results}

In total, twelve flares were identified in the light curves over the five-night observing run.  
All flares were detected in the $g$ and $r$ bands.  
Only three flares, observed on February 17, include $u$~band data, as $u$~band observations were conducted only on the first night before the filter malfunctioned.  
In addition, only three flares across all nights exhibited detectable amplitudes in the $i$ band.  
No flares were detected in the $z$ band.

\subsection{Flare energy}
For each flare, we computed the energy in the $u$, $g$, $r$, and $i$~bands, where data were available.  
A full table of energy measurements, along with other flare parameters including duration and peak amplitude, is provided in Table~\ref{tab:flare_properties}.  
The uncertainties in the flare energy estimates are primarily driven by the uncertainties in the equivalent duration measurements and the quiescent luminosity levels.



The most energetic flare in our sample is F4-0218, with energies of $1.17 \times 10^{30}$~erg in the $g$ band and $1.51 \times 10^{30}$~erg in the $r$ band.  
In contrast, the weakest flare observed, F9-0221, released $3.09 \times 10^{28}$~erg in the $g$ band and $6.53 \times 10^{28}$~erg in the $r$ band.  
All flares detected in the $u$ band exceeded $1 \times 10^{29}$~erg, and in each case, the $u$~band energy was greater than the corresponding energies in the $g$ and $r$ bands.

In general, flare energies in the $r$ band are consistently higher than those in the $g$ band, likely due to the broader passband and higher throughput in the red.  
We also find that $r$~band and $g$~band flare energies are strongly correlated, with a Pearson correlation coefficient of 0.95.  
Correlations between the $g$~band energies and those in the $u$ or $i$ bands could not be rigorously evaluated due to the limited number of detections in those bands.

\subsection{Flare temperature and filling factor}
\label{subsec:flare_temperature_discussion}

We estimated the flare temperatures for all detected events on an exposure-by-exposure basis using the $g$/$r$ flux density ratio.
For flares with detectable amplitudes in the $i$~band (i.e. F5-0220, F10-0221, and F12-0222), temperatures were also derived by SED fitting.  
We determined two temperature metrics for each flare:  
(1) the FWHM temperature ($T_{\mathrm{fl},\mathrm{fwhm}}$), defined as the average temperature within the full width at half maximum of the $g$~band flare light curve; and  
{(2) the global flare temperature ($T_{\mathrm{fl},\mathrm{glob}}$), defined as a single representative blackbody temperature characterizing the average energy per photon emitted during the entire flare. 
The FWHM temperature used here follows the same definition as the "peak flare temperature" in \citet{2020ApJ...902..115H}. 
We determine the FWHM from the $g$~band light curve because its flare profile is less affected by noise than the $r$-band profile.
While $T_{\mathrm{fl},\mathrm{fwhm}}$ may slightly underestimate the instantaneous peak temperature, it helps avoid the influence of single-epoch noise and serves as a physically meaningful proxy for the true peak flare temperature.
We use the term "FWHM flare temperature" to avoid confusion with the instantaneous true peak temperature.
On the other hand, the global flare temperature ($T_{\mathrm{fl},\mathrm{glob}}$) summarizes the flare’s spectral hardness and is especially useful when time-resolved temperature measurements are limited by signal-to-noise or observation cadence \citep[][]{2020PASJ...72...68N, 2020ApJ...902..115H}.}
In our case, the global temperature was estimated using the flare energy ratio between the $g$ and $r$~bands.

Using the derived temperatures, we also estimated the flare filling factor ($X_{\mathrm{BB}}$) based on Equation~\ref{eq:flare_temperature}.  
The temporal evolutions of flare temperature and filling factor across the flare phases for all detected events are shown in Figure~\ref{fig:flare_temperature_profiles}.
The peak and FWHM flare temperatures, along with the corresponding peak filling factors, are summarized in Table~\ref{tab:flare_temp_xbb}.

The hottest flare in our sample is F2-0217, with the temperatures of  $T_{fl,\mathrm{fwhm}} = 9900 \pm 1900$~K and $T_{fl,\mathrm{glob}} = 7400 \pm 300$~K, and a corresponding peak filling factor of $X_{\mathrm{BB}} = 8.35$~ppm.  
This is the only flare in our observations that reaches the empirical temperature of 10000~K.  
On the other hand, the coolest flare is F9-0221, with $T_{fl,\mathrm{fwhm}} = 3900 \pm 300$~K and $T_{fl,\mathrm{glob}} = 3400 \pm 200$~K, and a peak filling factor of $X_{\mathrm{BB}} = 77.24$~ppm.  
These results already demonstrate the significant variability in both flare temperature and emitting area on the same M~dwarf.  

On average, the flares in our study exhibit cooler temperatures than the canonical 10000~K value:  
$T_{fl,\mathrm{fwhm}} = 5500 \pm 1600$~K and $T_{fl,\mathrm{glob}} = 5600 \pm 1400$~K.  
These results follow the same general trend observed in other M dwarf flares, in which temperatures can be cooler than the empirical 10000~K, as also seen in flares from TRAPPIST-1 \citep{Maas+2022, 2023ApJ...959...64H}.  
The mean peak filling factor across our sample is $X_{\mathrm{BB}} = 151 \pm 119$~ppm.


\section{Discussion}
\label{sec:discussion}

\subsection{Balmer and blackbody continuum diagnostics from simultaneous broadband observations}
The flares F1-0217, F2-0217, and F3-0217 were observed in five different optical bands, including $u$~band, simultanousely. 
We noticed that the duration of flare detected in the $u$~band are much longer than those detected in other redder bandpasses (see Table~\ref{tab:flare_properties} and Figure~\ref{fig:Feb17_flares}).
This is likely because the emission enhancement within the $u$~band wavelength range is dominated by Balmer continuum emission from optically thin hydrogen recombination, rather than by the relatively optically thick hot blackbody component \citep[e.g.,][]{2013ApJS..207...15K}.  
The Balmer continuum typically exhibits a longer decay timescale than the hot blackbody emission \citep[][]{2016ApJ...820...95K}, which may explain the longer duration observed in the $u$~band.


The observed $u$~band peak flare amplitudes are lower than expected based on the derived flare temperatures, which contradicts the anticipated trend.  
For example, in the case of flare F2-0217, the peak temperature estimated from the $g$/$r$ flux ratio is {10156$\pm$476~K}, which predicts a $u$-band amplitude of 2.56.  
However, the observed peak amplitude is only 1.5.

Higher flare temperatures should correspond to stronger $u$~band emission, and the additional contribution from Balmer continuum is expected to make the total $u$~band flux even higher than that predicted by a blackbody extrapolation alone \citep[e.g.,][]{2006ApJ...644..484A}.
This discrepancy is likely due to observational bias introduced by the relatively long cadence ($\sim$6 minutes) used in our $u$~band observations.  
Long cadences are known to systematically underestimate peak flare amplitudes, especially for short-duration events \citep[e.g.,][]{2018ApJ...859...87Y, 2022AJ....163..164L, 2024AJ....168..234L}.

We simulated how the 360-second cadence underestimates flare amplitudes and derived a correction factor for the observed values.  
Synthetic flare profiles were generated at 0.1-second cadence using the analytic model from \citet{2017SoPh..292...77G}, with a 24-minute duration matching that of F2-0217.  
To mimic 6-minute cadence observations, we applied a rolling average with a randomly shifted 6-minute window, and we repeated this process 1000 times.
From the 1000 6-minute synthetic light curve, we found that, on average, the 6-minute cadence underestimated the peak flare amplitude by 60\%, meaning the observed amplitude represents only about 40\% of the actual value.
For F3-0217, whose duration $\approx~36$~minutes, the flare amplitude observed at 6-minute cadence is expected to be 47\% of the actual value.
Flare F1-0217 was excluded from the analysis because the impulsive and peak phases were not captured in our observations.

The corrected $u$-band peak amplitude of flare F2-0217 is $3.75 \pm 0.38$, which exceeds the value predicted by a blackbody model with $T_{\mathrm{fl}} = 10156$~K. 
We attribute this excess to additional Balmer continuum emission. 
To explore this further, we estimated the Balmer jump ratio ($R_{\mathrm{BaJ}}$), similar to the C3615/C4170 ratio defined by \citet{2013ApJS..207...15K}, using the broadband SED at flare peak. We note that our calculation differs slightly, as the effective central wavelength of our $u$~band filter is 3540~\AA, compared to the 3615~\AA\ reference used in their definition.

We measured \( R_{\mathrm{BaJ}} = 1.76 \pm 0.23 \) and \( C4170/C6010 = 2.06 \pm 0.31 \) of F2-0217, placing F2-0217 within the empirical flare color-color distribution shown in Figure~9 of \citet{2019ApJ...871..167K}, and consistent with the impulsive-type (IF) regime.
In this context, the flare continuum is likely dominated by hot blackbody-like radiation, produced by heating at high column mass, as seen in RHD models with strong electron beam fluxes \citep[e.g., F13,][]{2016ApJ...820...95K}. 

Flare F3-0217 shows a larger corrected $u$-band amplitude of $5.08 \pm 0.38$, higher than the $u$~band amplitude of $\approx 2.16$ derived from its peak $T_{\mathrm{fl}} = 6351$~K.
Nevertheless, the Balmer Jump ratio and Blue-to-red continuum ratio are \( R_{\mathrm{BaJ}} = 2.07\pm0.15 \) and \( C4170/C6010 = 1.16\pm0.13 \), respectively, placing it outside the main trend of M dwarf flares in the K19 diagram. 
Although the Balmer jump is low enough to be consistent with IF flares, the redder optical slope and low inferred blackbody temperature of $\sim$6300~K are more typical for gradual (GF) or hybrid (HF) events.
This may result from spatial or temporal averaging over multiple flare kernels, each contributing blackbody-like continuum components with varying temperatures. 
Simultaneous spectroscopic observations covering the $u$~band wavelength range will be necessary to confirm the presence and relative strength of continuum components.
Additionally, detailed radiative-hydrodynamic (RHD) modeling will be critical to interpret the underlying atmospheric conditions responsible for this flare’s unique continuum emission.

\subsection{Flare temperatures and energies}
\label{subsect:flare_temperature_and_energy}
\citet{2020ApJ...902..115H} demonstrated a positive correlation between flare energy and blackbody temperature in M dwarfs, based primarily on a sample of superflare events with bolometric energies $\geq 10^{32}$~erg.  
In this study, we extend and refine that correlation toward the lower-energy regime using flares from the late-M dwarf Wolf~359, along with flare data from TRAPPIST-1 reported by \citet{Maas+2022} and \citet{2023ApJ...959...64H}.  



We converted the bolometric flare energies reported by \citet{Maas+2022}, the TESS~band energies reported by \citet{2023ApJ...959...64H}, {and Evryscope $g^{'}$~band energies reported by \citet{2020ApJ...902..115H} into TRIPOL} $g$~band energies using the approximation provided by \citet{2018ApJ...860L..30H}:
\begin{equation}
\label{eq:energy_conversion}
    E_{\mathrm{fl, bol}} = \frac{E_{\mathrm{fl},\lambda}}{f_{\lambda}(T_{\mathrm{fl}})},
\end{equation}
where $E_{\mathrm{fl, bol}}$ is the bolometric flare energy, $E_{\mathrm{fl},\lambda}$ is the flare energy in the bandpass $\lambda$, and $f_{\lambda}(T_{\mathrm{fl}})$ is the bandpass response factor, which depends on the flare temperature $T_{\mathrm{fl}}$ and the spectral sensitivity of the bandpass $\lambda$.
The response factor functions for the TESS and {Evryscope $g^{'}$~bands can be found in Figure~3 of \citet{2020ApJ...902..115H}.}
{The response factor function of the TRIPOL $g$~band is shown in Figure~\ref{fig:tripol_response_factor_function}.}
{We note that, in this study, we include only the flare sample from M~dwarfs (\(M_{*} \leq 0.42\,M_{\odot}\)) from \citet{2020ApJ...902..115H} in our analysis and discussion.}


{Figure~\ref{fig:flare_temperature_vs_energy} shows the relationship between flare TRIPOL~$g$~band energies and FWHM temperatures}, incorporating data from this study as well as from previous studies.
{The data in this figure are summarized in Table~\ref{tab:m_dwarfs_flares_in_temperature_energy_figure}.}
We find that less energetic flares tend to exhibit lower temperatures compared to more energetic flares, although a few exceptions exist.  
This is expected, as the estimated flare energy depends not only on temperature but also on other properties such as duration \citep[e.g.,][]{2015ApJ...798...92W}.  


{We note that there is a gap in the flare energy between $E_{fl,g} \sim 10^{30} - 10^{32}$~erg, which will require further monitoring and observations to fill in future work.}
{To avoid bias in the fitting results caused by this data gap, we determined the power-law relations between FWHM flare temperatures and flare energies by using only the flares in the low-energy regime (\(E_{\mathrm{fl},g} < 10^{31}~\mathrm{erg}\)). 
We then extrapolated the fitted relation to the high-energy regime to enable comparison with results from \citet{2020ApJ...902..115H}.}
{In addition, we applied the same power-law relation determination to the sample of field M~dwarf superflares reported by \citet{2020ApJ...902..115H}, and then extrapolated the superflare-derived relation down to the solar-class flare regime to enable direct comparison with our results.}

{We employed the Bayesian Markov Chain Monte Carlo framework implemented in the Python package \texttt{pymc3} to derive the best-fit coefficients and their uncertainties in a statistically robust manner.
{The resulting power-law coefficients ($\alpha$, $\beta$) describing the correlation between FWHM flare temperature and flare energy in the TRIPOL $g$ band are based on the relation}
\begin{equation}
    \label{eq:flare_energy_temperature}
    \log(T_{\mathrm{fl},\mathrm{fwhm}}) = \alpha\, \log(E_{\mathrm{fl},\lambda}) + \beta,
\end{equation}
{  
The coefficients derived separately for superflares and for solar-class flares are summarized in Table~\ref{tab:flare_temperature-vs-energy_powerlaw}, and the corresponding relations are illustrated in Figure~\ref{fig:flare_temperature_vs_energy}.
We also list the power-law coefficients determined by \citet{2020ApJ...902..115H} for superflares in EvryScope~$g'$~band in Table~\ref{tab:flare_temperature-vs-energy_powerlaw} for comparison.}

{As shown in Figure~\ref{fig:flare_temperature_vs_energy}, it is visually obvious that the extrapolation of the power-law relations derived from M~dwarf superflares by \citet{2020ApJ...902..115H} to lower energies may underestimate the temperatures of solar-class flares, regardless of the flare energy bands.} 
{Nevertheless, despite the lack of data points in the intermediate energy range (\(E_{\mathrm{fl},g} \sim 10^{30} - 10^{32}~\mathrm{erg}\)), the power-law relations we derived for solar-class flares in this study, together with those from \citet{Maas+2022} and \citet{2023ApJ...959...64H}, are consistent, within the uncertainties, with the high-energy superflares from field M~dwarfs reported by \citet{2020ApJ...902..115H}.}
We evaluated the root mean square error (RMSE) between the best-fit power-law and the observed flare temperatures across solar-class flares and superflares regimes.  
{For flares with \(E_{\mathrm{fl,g}} < 10^{31}~\mathrm{erg}\), the RMSE is approximately 600~K, whereas for flares with \(E_{\mathrm{fl,g}} \geq 10^{31}~\mathrm{erg}\), the RMSE increases to about 7600~K.  
This suggests that the temperatures of less energetic, solar-class flares can be more accurately predicted by the power-law relation compared to those of higher-energy superflares.
The Pearson correlation coefficient for the \(\log(T_{\mathrm{fl},\mathrm{fwhm}})\)--\(\log(E_{\mathrm{fl},g})\) relation across all energy regimes is about 0.66, indicating a moderate positive correlation between these two properties.}
This strengthens our confidence in the power-law relations derived from the low-energy regime and supports their apparent agreement with the high-energy flare population when extrapolated.
This agreement, while not guaranteed due to the extrapolative nature of the fit, suggests that the underlying physical processes governing flare heating may scale similarly across several orders of magnitude in energy.
Future observations that fill the intermediate energy range will be essential to validate this continuity.
On the other hand, the less accurate predictions made by the power-law fit for the superflares may indicate that such energetic events involve more complex physical mechanisms that cannot be fully captured by a single blackbody model as a simplified approximation.

\subsection{Flare temperature frequency distribution}
\label{subsec:flare_temperature_frequency_distribution_and_PAR}
%
We investigated the flare temperature frequency distribution for an M~dwarf for the first time.  
Figure~\ref{fig:flare_temperature_frequency_distribution} presents the cumulative frequency distributions of FWHM flare temperatures ($T_{\mathrm{fl},\mathrm{fwhm}}$) and global flare temperatures ($T_{\mathrm{fl},\mathrm{glob}}$) for Wolf~359.  
Similar to flare energy distributions, we expressed the cumulative temperature distributions as linear functions in logarithmic form.  
{The power-law relations for the two temperature metrics, determined using \texttt{pymc3}, are given by
\begin{equation}
    \label{eq:fwhm_temperature_distribution}
    \log(N_{fl}) = ( -2.92 \pm 0.31) \log(T_{\mathrm{fl},\mathrm{fwhm}}) + (13.81 \pm  1.16),
\end{equation}
\begin{equation}
    \label{eq:global_temperature_distribution}
    \log(N_{fl}) = (-2.87 \pm 0.21) \log(T_{\mathrm{fl},\mathrm{glob}}) + (13.78 \pm  0.79),
\end{equation}
where $N_{fl}$ is the cumulative frequency of flares in units of flare number $\mathrm{yr}^{-1}$.}

{A power-law slope of $-1.13 \pm 0.14$ has been reported for the cumulative flare frequency distribution in terms of flare energy for Wolf~359 based on 872 flares detected using K2 data \citep{2021AJ....162...11L}. 
{Given that the bolometric flare energy–FWHM temperature relation determined solely from solar-class flares (see Equation~\ref{eq:flare_energy_temperature} and Table~\ref{tab:flare_temperature-vs-energy_powerlaw}) appears to hold across a broader energy range, we expect a corresponding FWHM flare temperature distribution slope of {$\sim -10.9$} when combined with the result from \citet{2021AJ....162...11L}.}
This is significantly steeper than our measured slope of $-2.92 \pm 0.26$ in Equation~\ref{eq:fwhm_temperature_distribution}.}
{This discrepancy is likely due to the limited size and low-energy feature of our flare sample.
Such a small sample, particularly in the low-energy regime, is susceptible to detection incompleteness—especially for short-duration, low-amplitude flares—which tends to bias the slope toward shallower values \citep[e.g.,][]{2023A+A...669A..15Y, 2024AJ....168..234L}. 
The slope of {$\sim -10.9$} inferred from the energy distribution and temperature–energy relation may thus represent a more accurate expectation, which future observations with larger and more complete flare samples will be able to test and confirm.}




\subsection{{Influence of stellar flares on photosynthesis of hypothetical Earth-analogues}}
Photosynthesis is the dominant basis of life on Earth and is believed to have arisen early in Earth's history \citep[e.g.,][]{2016AREPS..44..647F}.  
As a result, photosynthetically active radiation (PAR; 400–750~nm), which spans the spectral range most effective at driving photosynthesis, is considered a key requirement for sustaining Earth-like biospheres on habitable planets {\citep{1998Sci...281..237F, 2018ApJ...859..171L, 2018ApJ...865..101M}.}

To assess whether stellar flares could help compensate for the PAR deficit around low-mass stars, \citet{2019MNRAS.485.5924L} estimated the additional PAR contribution from flares by modeling them as blackbodies with a fixed temperature of 9000~K.  
They concluded that planets orbiting low-mass stars, particularly M~dwarfs with masses $\lesssim 0.2~M_\odot$, receive significantly less quiescent PAR than Earth, potentially limiting their ability to support net primary productivity (NPP) or accumulate detectable levels of atmospheric oxygen.


{In this study, we evaluate the flare-contributed PAR to a hypothetical Earth-like planet in Wolf~359's habitable zone by combining our derived flare temperature--bolometric energy correlation with the flare energy frequency distribution reported by \citet{2021AJ....162...11L}, providing a more realistic assessment than assuming a single fixed temperature.}
We first modified the equation for the minimum occurrence rate of flares, $ {{\nu}}_{\mathrm{fl}}$, with bolometric energy $E_{\mathrm{fl,bol}}$ that could contribute additional PAR sufficient to sustain Earth-like NPP on this hypothetical planet, starting from Equation~6 in \citet{2019MNRAS.485.5924L}
{
\begin{equation}
\label{eq:Lingam}
2.6\times10^{18}~ {{\nu}}_{\mathrm{fl}} ~\frac{\epsilon_{\mathrm{PAR}}(T_{fl}) E_{\mathrm{fl, bol}}(T_{fl}) }{4 \pi a_{\star}^{2}} \gtrsim F_{\oplus}.
\end{equation}
}
In this equation, $F_{\oplus} = 4 \times 10^{20}$~photons~m$^{-2}$~s$^{-1}$ represents the critical PAR photon flux incident on a planet that is required to sustain a net primary productivity equivalent to that of Earth.  
$\epsilon_{\mathrm{PAR}}(T_{\mathrm{fl}})$ is the fraction of the total flare energy emitted within the PAR wavelength range, expressed as a function of flare temperature (Figure~\ref{fig:epsilon_PAR}).  
$E_{\mathrm{fl, bol}}(T_{\mathrm{fl}})$ is the bolometric flare energy as a function of flare temperature, which can be derived from Equation~\ref{eq:flare_energy_temperature} with the coefficients from solar-class flares in Table~\ref{tab:flare_temperature-vs-energy_powerlaw}.  
$a_{\star} = 0.0377$~AU is the semimajor axis at which an Earth-like planet would receive the same effective surface temperature in the Wolf~359 system.
As a result, the equation can be expressed as
\begin{equation}
    \nu_{fl} \gtrsim 1.35\times10^{6}~yr^{-1}~(\frac{\epsilon_{\mathrm{PAR}}(T_{fl})~E_{fl,bol}(T_{fl})}{10^{34}~erg})^{-1}(\frac{R_{*}}{R_{\odot}})^{2} (\frac{T_{*}}{T_{\odot}})^{4},
\end{equation}
where $R_{*}=0.16R_{\odot}$ and $T_{*}=2800$~K for Wolf~359 \citep[][]{stellar_radius, 1991ApJS...77..417K}; $T_{\odot}=5800$~K.

{We adopt the FWHM flare temperature–TRIPOL~$g$ energy correlation (Equation~\ref{eq:flare_energy_temperature}, using the solar-class flare coefficients in Table~\ref{tab:flare_temperature-vs-energy_powerlaw}) together with the flare energy frequency distribution from \citet{2021AJ....162...11L} (see their Equation 10 and Section 4.3). The bolometric flare energies are then estimated from the TRIPOL~$g$ energies using the conversion relation in Equation~\ref{eq:energy_conversion} and the response factor function of TRIPOL~$g$ in Figure~\ref{fig:tripol_response_factor_function}.
The PAR efficiency $\epsilon_{\mathrm{PAR}}$ reaches its peak value of 0.44 at $T_{\mathrm{fl},\mathrm{fwhm}} \approx 6800$~K, corresponding to a bolometric flare energy of {$E_{\mathrm{fl,bol}} \sim 9 \times 10^{31}$~erg.}
The minimum occurrence rate of such flare required to sustain Earth-like NPP is {$\nu_{\mathrm{fl}} = 4.7 \times 10^{5}$~yr$^{-1}$ in the Wolf~359 system.}}
{In comparison, the cumulative frequency of flares with $E_{\mathrm{fl,bol}} \sim 9 \times 10^{31}$~erg. observed in Wolf~359 is $N_{\mathrm{fl}}(E_{\mathrm{fl,bol}} \geq 9 \times 10^{31}~erg) \approx 176$~yr$^{-1}$, much less than the minimum flare rate requirement.}
This suggests that typical superflares are insufficient to sustain Earth-like net primary productivity on habitable-zone exoplanets, even in extremely active systems like Wolf~359.

{At the high-energy end, M~dwarf flares can reach energies as high as {$1\times10^{36}$~erg \citep[][]{2020ApJ...902..115H, 2024AJ....168..234L}, corresponding to an FWHM flare temperature of approximately $T_{\mathrm{fl},\mathrm{fwhm}} = 16500$~K.}}
For such energetic flares, the minimum Earth-like-NPP-sustainable flare occurrence rate is {$\nu_{\mathrm{fl}} = 60$~yr$^{-1}$.}
{Using the power-law energy frequency distribution \cite{2021AJ....162...11L} derived for Wolf~359, the predicted cumulative flare frequency of $N_{\mathrm{fl}}(E_{\mathrm{fl,bol}} \geq 10^{36}~erg) \approx 4\times10^{-3}$~yr$^{-1}$.}
Under these extreme flare conditions, the flare activity of Wolf~359 is still far from satisfying the PAR threshold necessary to sustain Earth-like photosynthesis on a habitable-zone planet.}

\section{Conclusions}
\label{sec:conclusions}
We presented simultaneous multiband optical observations of flares on Wolf~359, obtained with the Lulin 1-m and 41-cm telescopes, to investigate the flare temperature distribution and its implications for stellar and planetary environments.  
Over the course of five nights, we detected a total of twelve flares.  
Flare temperatures were determined using the $g$/$r$ flux density ratio, and, when detectable in the $z$~band, through spectral energy distribution (SED) fitting.  
We introduced two types of flare temperatures: (1) the FWHM flare temperature, $T_{\mathrm{fl,fwhm}}$, and (2) the global flare temperature, $T_{\mathrm{fl,glob}}$, as defined in Section~\ref{subsec:flare_temperature_discussion}.  
The average FWHM flare temperature is $T_{\mathrm{fl,fwhm}} = 5500 \pm 1600$~K, and the average global flare temperature is $T_{\mathrm{fl,glob}} = 5600 \pm 1400$~K, both significantly cooler than the canonical value of 10000~K.


In addition to evaluating flare temperatures, we investigated the wavelength-dependent evolution of flares F1-0217, F2-0217, and F3-0217.  
We observed that $u$~band flare durations were notably longer than in redder bands, likely due to optically thin Balmer continuum emission with slower decay timescales.  
Observed $u$~band peak amplitudes were lower than predicted by flare temperatures, but cadence simulations show that long exposures ($\sim$6 minutes) can significantly underestimate peak amplitudes, particularly for impulsive flares.  
After correction, the $u$~band fluxes exceed blackbody model predictions, consistent with additional Balmer continuum emission.  
Flare F2-0217 shows properties consistent with impulsive-type events, dominated by hot blackbody-like radiation, while F3-0217 exhibits a redder continuum and a lower inferred blackbody temperature, suggesting a more complex or spatially averaged structure.  
These findings highlight the importance of cadence corrections in photometric flare studies and demonstrate the value of simultaneous multiband observations for disentangling flare continuum components.  
Future high-cadence spectroscopic monitoring will be crucial for refining models of continuum evolution and understanding the diverse heating processes in Wolf~359's flares.


{We further investigated the correlation between {FWHM} flare temperature and {TRIPOL~$g$} energy by extending the sample toward lower-energy flares, incorporating data from Wolf~359, TRAPPIST-1, and previous M~dwarf studies \citep{Maas+2022, 2020ApJ...902..115H, 2023ApJ...959...64H}. 
To enable a consistent comparison, we converted all reported flare energies into TRIPOL \(g\)~band energies using response factor functions that account for flare temperature and instrument sensitivity.
We determined power-law relations between FWHM flare temperature and both \(g\)~band in the low-energy regime (\(E_{\mathrm{fl},g} < 10^{31}~\mathrm{erg}\)) and extrapolated these relations to higher energies.
As a result, {despite the lack of data points in the energy range (i.e., \(E_{\mathrm{fl},g} \sim 10^{30} - 10^{32}~\mathrm{erg}\)), the power-law relations we derived for solar-class flares agree, within the error bars, with the M~dwarfs' superflares by \citet{2020ApJ...902..115H}.
The Pearson correlation coefficients of approximately 0.66 for \(\log(T_{\mathrm{fl},\mathrm{fwhm}})\)--\(\log(E_{\mathrm{fl},g})\) relation across all energy regimes further support a moderate positive correlation between flare energy and temperature.}
While this agreement is not guaranteed—given the extrapolative nature of the fit and the limited size of our flare sample—it suggests that the physical processes responsible for flare heating may scale similarly from weak to energetic flares. 
However, superflares' temperatures exhibit greater scatter compared to those of the lower-energy, solar-class flares. 
This reduced consistency observed in the higher-energy regime may hint that superflares involve more complex or diverse physical mechanisms than those operating in solar-class flares, potentially limiting the applicability of simple blackbody-based models at extreme energies.
Future observations that fill in the intermediate energy regime will be critical for confirming this apparent continuity and refining the underlying scaling relations.}

{We derived, for the first time, the power-law slope of the flare temperature frequency distribution for the M~dwarf Wolf~359.
However, due to the limited sample size, the empirically derived slope may be biased and appears shallower than expected.
{This is supported by} our FWHM flare temperature–energy relation and the flare energy frequency distribution slope reported by \citet{2021AJ....162...11L}.
A more complete and statistically robust flare dataset will be essential in the future to more accurately characterize the flare temperature frequency distribution.}

We also explored the broader implications of our findings for planetary habitability of hpyerthetical HZ Earth-like planet around Wolf~359.
{Using our derived flare temperature–-TRIPOL~$g$ energy correlation, bolometric flare energies conversion, and the Wolf~359's flare energy frequency distribution reported by \cite{2021AJ....162...11L}, we assessed their potential contribution to PAR in the habitable zone.  
We find that typical solar-class giant flares ($E_{\mathrm{fl,bol}} \sim 9\times10^{31}$~erg, $T_{\mathrm{fl,fwhm}} \sim 6800$~K) are too infrequent to sustain Earth-like NPP.
At the high-energy end of superflare regime ($E_{\mathrm{fl,bol}} \sim 1\times10^{36}$~erg, $T_{\mathrm{fl,fwhm}} \sim 16500$~K), flare activity of Wolf~359 still cannot meet the PAR threshold.}

Finally, the frequent occurrence of cool flare temperatures suggests that contamination of NIR transmission spectra, due to stellar surface heterogeneity induced by flaring activity, may be more common than previously anticipated, potentially impacting the characterization of exoplanetary atmospheres \citep[][]{2018ApJ...853..122R, 2019AJ....157...96R, 2024AJ....168...82R, 2023ApJ...955L..22L, 2023ApJ...959...64H}.  
The flare temperature distribution derived in this study provides valuable input for better understanding and modeling these effects and could ultimately aid in mitigating flare activity contamination when interpreting exoplanet atmospheric compositions.







\begin{acknowledgments}
We thank the anonymous referee for their constructive and insightful comments, which have significantly improved the quality of this paper. This research was supported by grant No. 113-2112-M-008-002 from the National Science and Technology Council (NSTC) of Taiwan. The data analyzed in this study were obtained from the Lulin Observatory in Taiwan, which is operated and maintained by the Graduate Institute of Astronomy at National Central University.
\end{acknowledgments}

\clearpage

\begin{figure}
    \centering
    \figurenum{2}
    \epsscale{1}
    \plotone{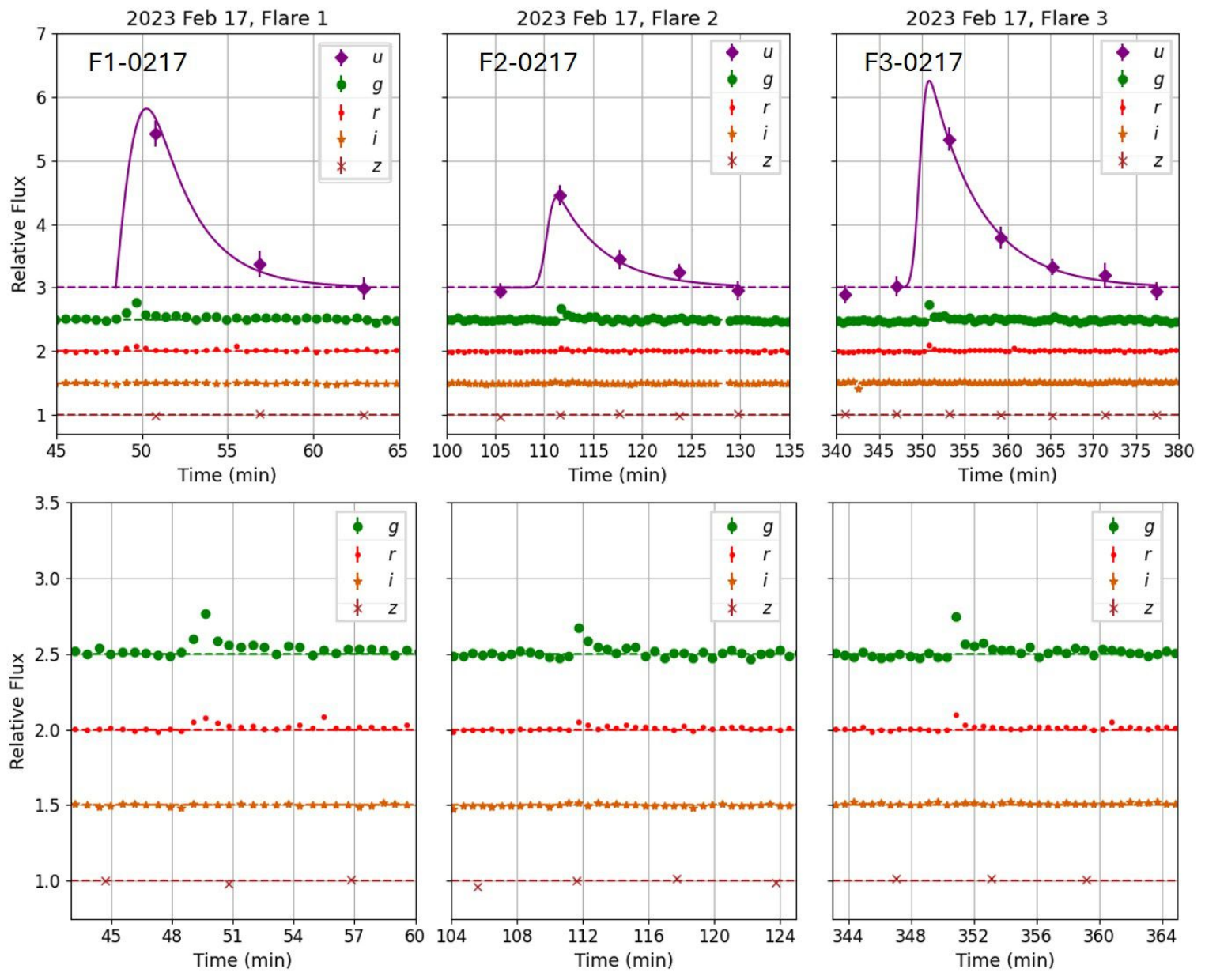}
    \caption{
    Flares detected in Wolf~359 on 2023 February 17 in multiple photometric bands simultaneously.  
    \textbf{Top row:} Light curves in the $u$, $g$, $r$, $i$, and $z$ bands for each of the three detected flares.  
    The purple solid curves represent synthetic flare profiles constructed to approximate the full flare variation in the $u$~band based on the observed data points.
    {These profiles were determined by using the flare profile model from \cite{2017SoPh..292...77G}, specifically their Equations 1 through 4.}
    \textbf{Bottom row:} Zoom-in profiles of the $g$, $r$, $i$, and $z$~band light curves.
    Dashed horizontal lines indicate the quiescent flux levels in each band.
    The flares exhibit significant brightness enhancements in the $u$, $g$, and $r$ bands, with much weaker or undetectable signals in the $i$ and $z$ bands.  
    }
    \label{fig:Feb17_flares}
\end{figure}

\begin{figure}
    \centering
    \figurenum{2}
    \epsscale{0.9}
    \plotone{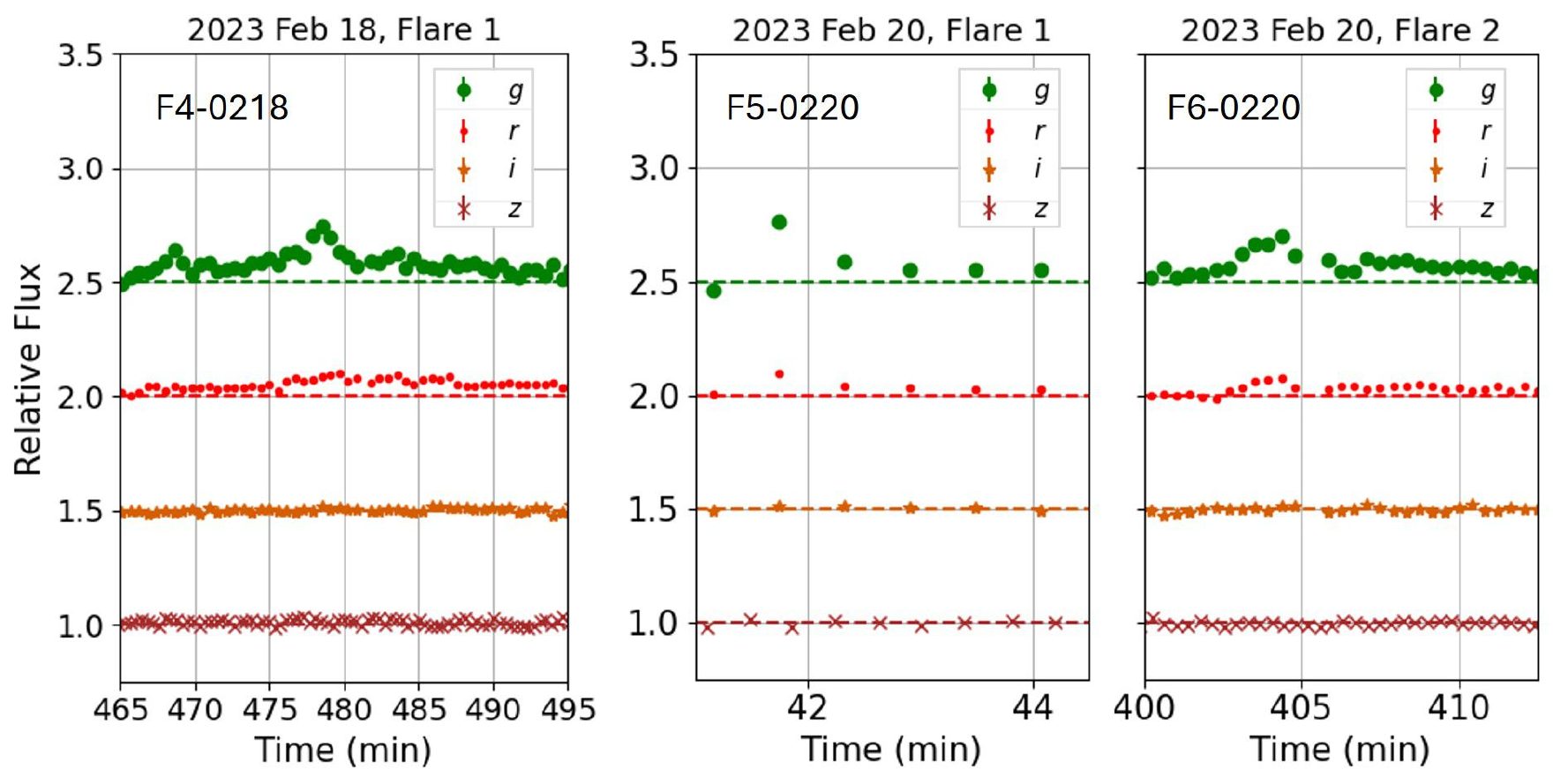}
    \caption{(continued): Flares detected on 2023 February 18 and 20. 
    Starting from February 18, observations were conducted only in the $g$, $r$, $i$, and $z$~bands due to a malfunction of the $u$~band filter.}
    \label{fig:Feb18_20_flares}
\end{figure}

\begin{figure}
    \centering
    \figurenum{2}
    \epsscale{1}
    \plotone{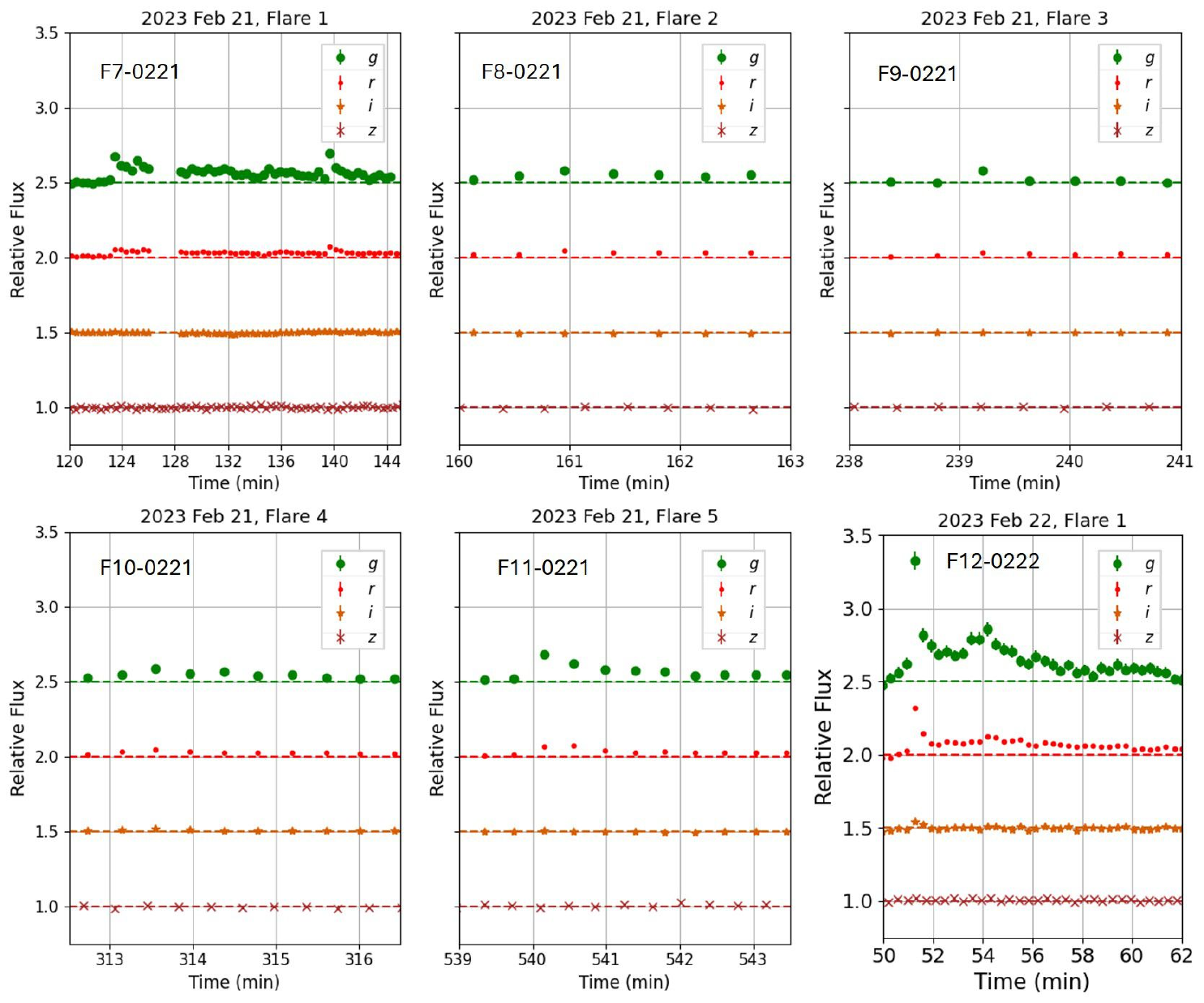}
    \caption{(continued): Flares detected on 2023 February 21 and 22.
    In total, twelve flares were observed in Wolf~359 during our observation campaign.}
    \label{fig:Feb21_22_flares}
\end{figure}

\begin{figure}
    \centering
    \figurenum{3}
    \epsscale{0.7}
    \plotone{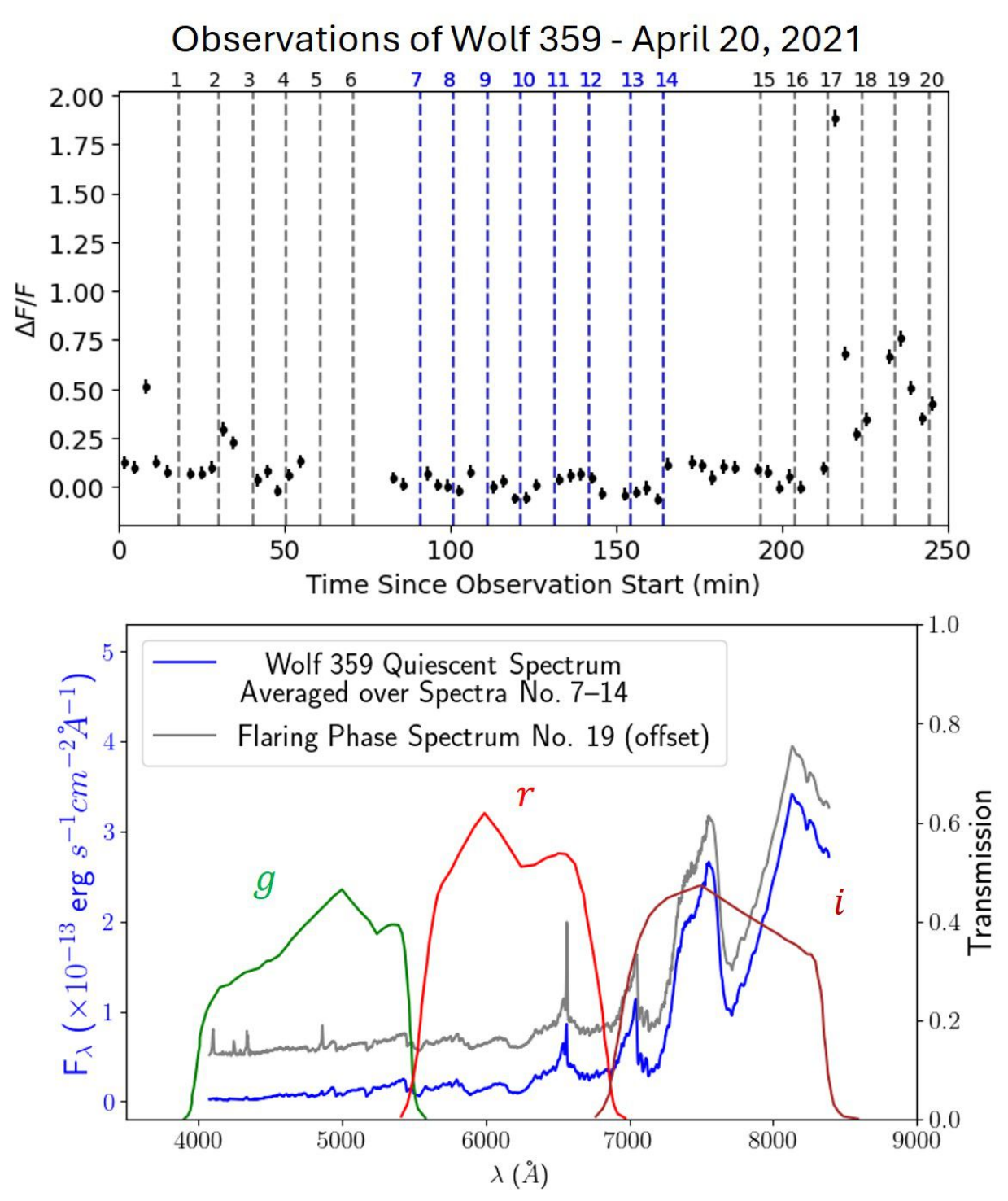}
    \caption{
    \textbf{Top:} $u$-band light curve of Wolf~359 obtained on 2021 April 20, showing relative flux variability throughout the observation. 
    Vertical dashed lines marks the midpoint of each spectroscopic exposure taken by using the Shelyak LISA low-resolution spectrograph ($R \sim 1000$) on the Lulin 1-m telescope. 
    Spectra numbered 7 through 14 (blue lines) were identified as taken during a flare-free interval and were used to construct the quiescent spectrum.
    \textbf{Bottom:} The quiescent spectrum of Wolf~359 (blue), obtained by averaging spectra No.~7--14. 
    The spectrum in gray represents the spectrum No.~19 observed during the flaring phase and is offset by 0.5 units for better visual comparison.
    The transmission curves of the $g$ (green), $r$ (red), and $i$ (brown) filters used by TRIPOL are also shown.
    This quiescent spectrum was used to calculate the star's quiescent luminosities in the $g$, $r$, and $i$ bands.
    }
    
    \label{fig:quiescent_spectrum_wolf359}
\end{figure}

\begin{figure}
    \centering
    \figurenum{4}
    \epsscale{1}
    \plotone{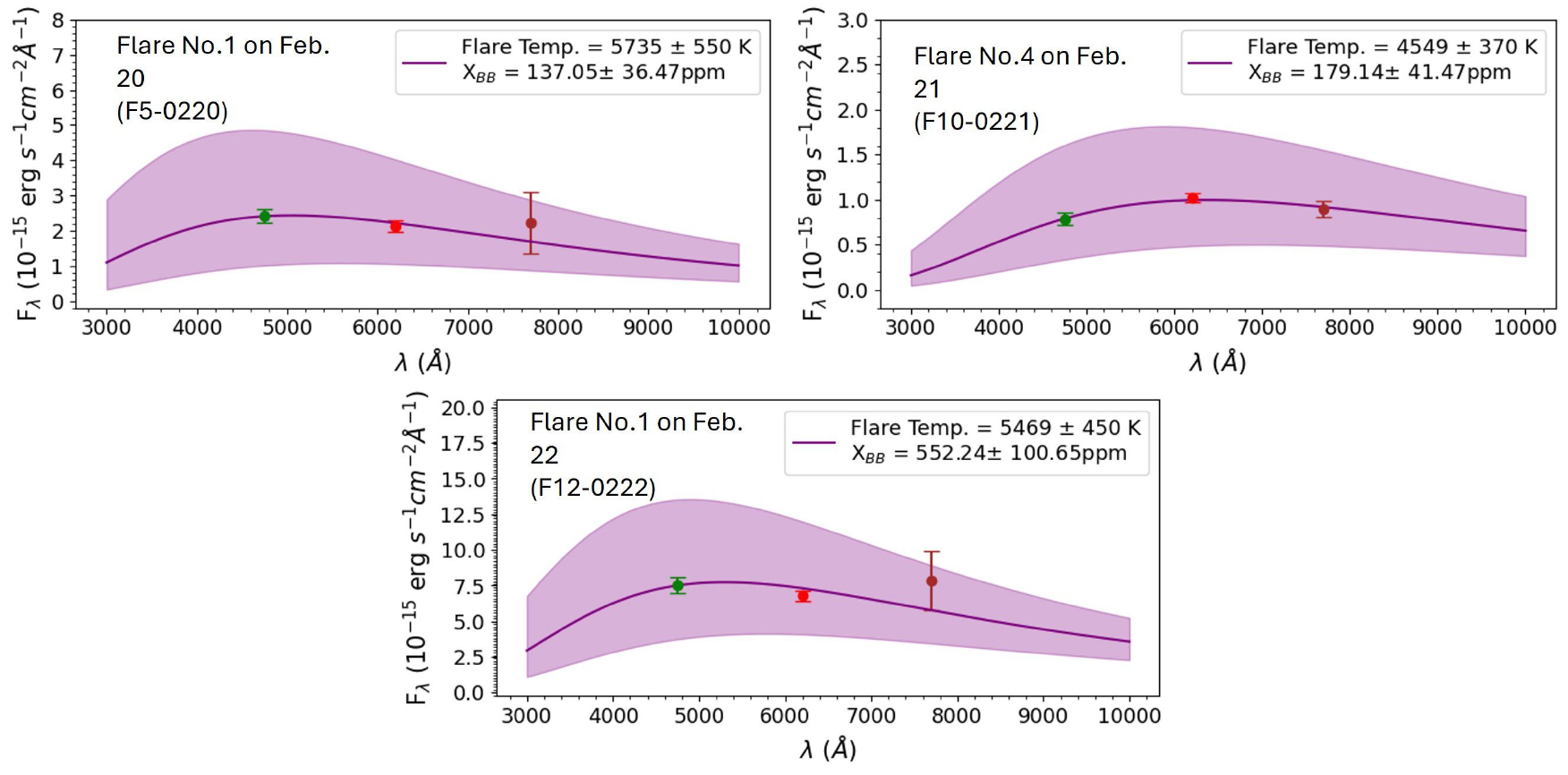}
    \caption{Peak flare-only SEDs and best-fit blackbody models for three flares detected in the $g$, $r$, and $i$ bands: F5-0220 (Flare No.~1 on February 20), F10-0221 (Flare No.~4 on February 21), and F12-0222 (Flare No.~1 on February 22).  
    Colored data points represent the flare-only flux densities derived for each band ($g$~band in green, $r$~band in red, and $i$~band in brown), and the solid purple lines show the corresponding blackbody fits, with shaded regions indicating 1--$\sigma$ uncertainties.  
    The derived peak flare temperatures are $T_{fl,\mathrm{SED}} = 5700 \pm 600$~K, $4500 \pm 400$~K, and $5500 \pm 500$~K, respectively.  
    }
    \label{fig:sed_fitting_examples}
\end{figure}

\begin{figure}
    \centering
    \figurenum{5}
    \epsscale{1}
    \plotone{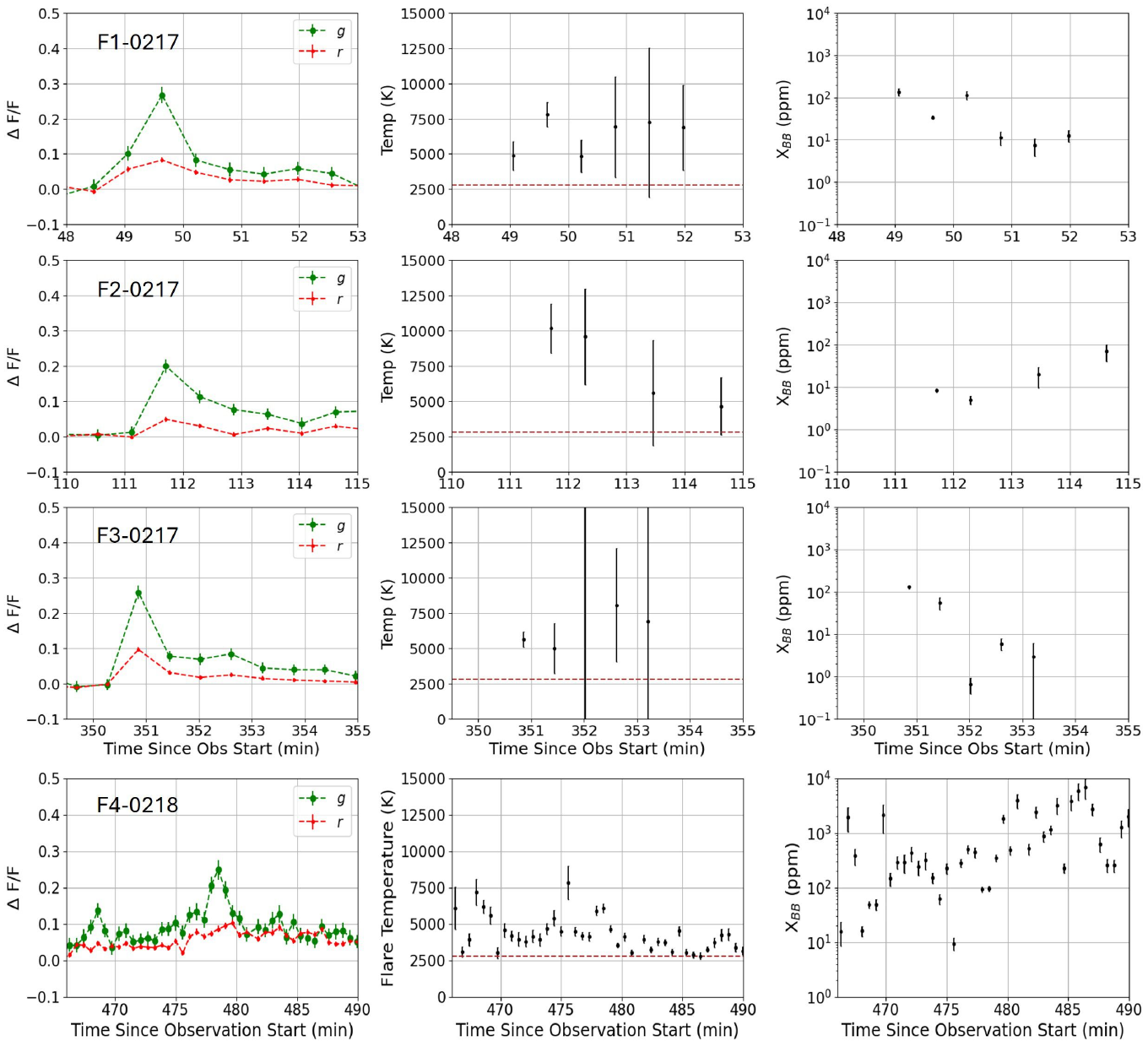}
    \caption{Temporal evolution of flare temperature and filling factor ($X_{\mathrm{BB}}$) for all flares analyzed in this study. Each row corresponds to an individual flare event. 
    {The left panels display the flare light curves in the $g$ (green) and $r$ (red) bands, with the flare identification labeled accordingly.
    The middle panels illustrate the flare temperature as a function of time, derived from the $g$/$r$ flux density ratio. The brown horizontal dashed line represents the effective temperature of Wolf~359, which is 2800~K.
    The right panels show the corresponding evolution of the filling factor ($X_{\mathrm{BB}}$) throughout the flare duration.}
    For flares {F5-0220, F10-0221, and F12-0222, which exhibited detectable amplitudes in the $i$ band, the $i$-band light curves are also shown.}}
    \label{fig:flare_temperature_profiles}
\end{figure}

\begin{figure}
    \centering
    \figurenum{5}
    \epsscale{1}
    \plotone{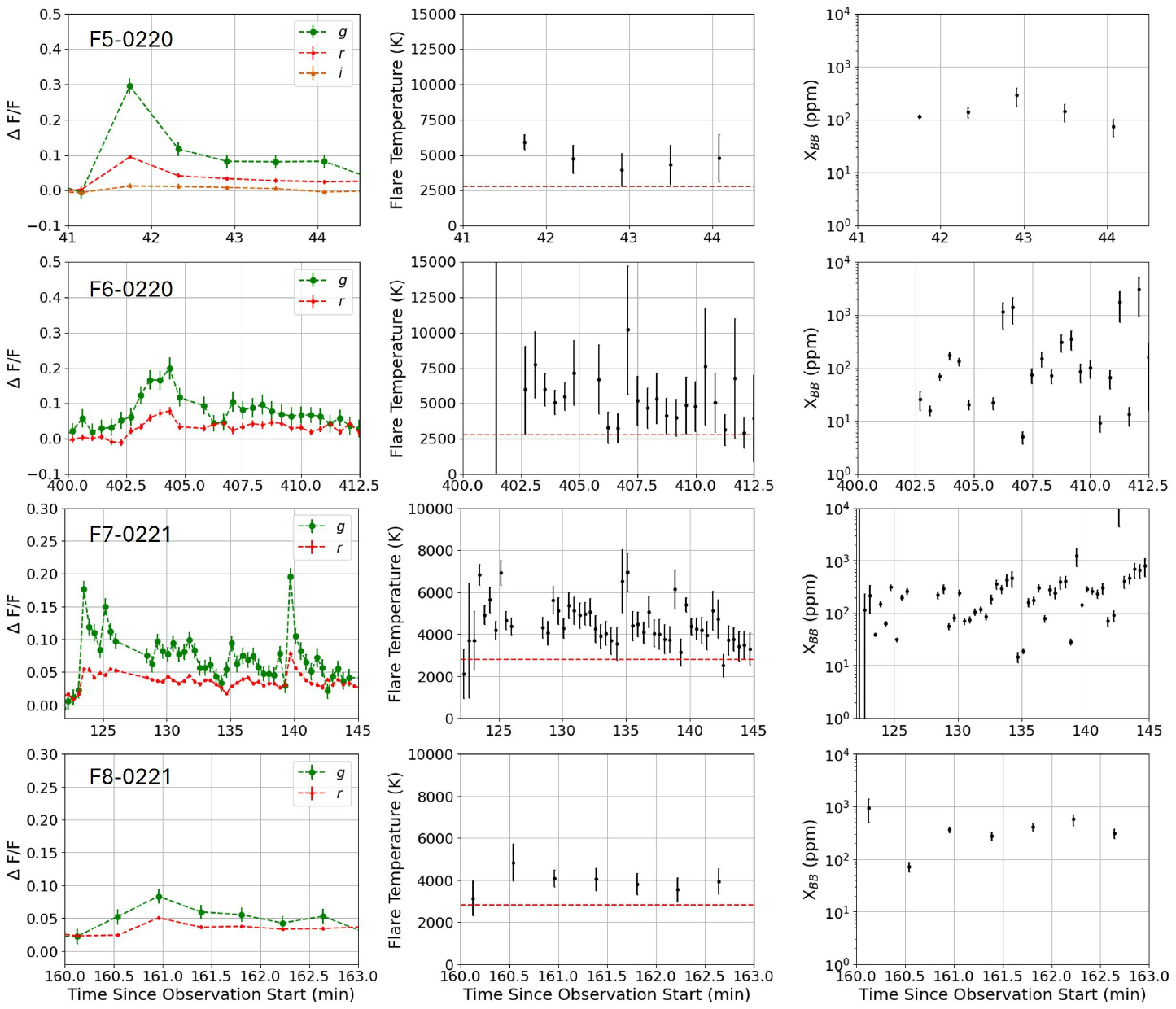}
    \caption{(continued)}
    \label{fig:flare_temperature_profiles}
\end{figure}

\begin{figure}
    \centering
    \figurenum{5}
    \epsscale{1}
    \plotone{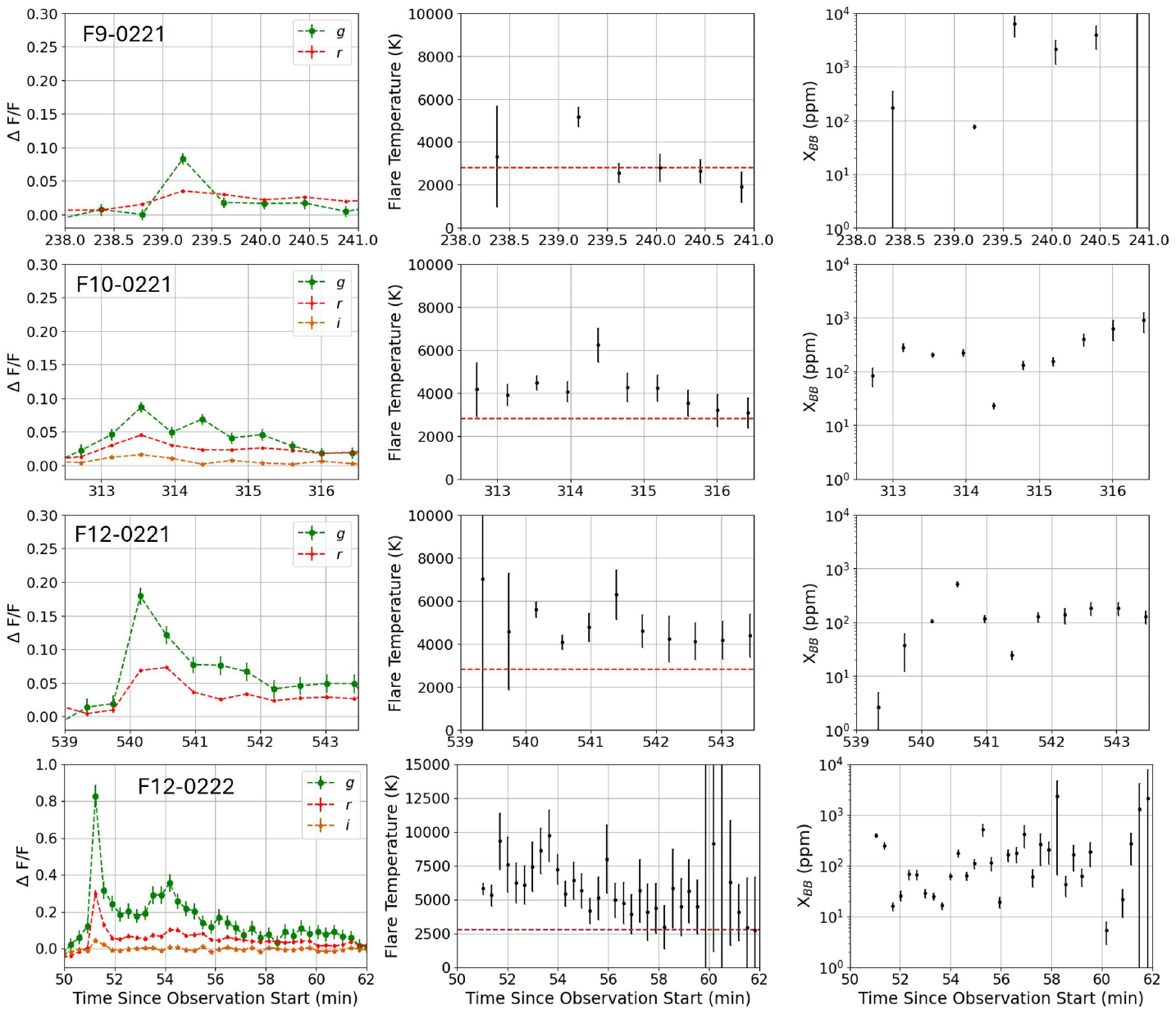}
    \caption{(continued)}
    \label{fig:flare_temperature_profiles}
\end{figure}


\begin{figure}
    \centering
    \figurenum{6}
    \epsscale{0.5}
    \plotone{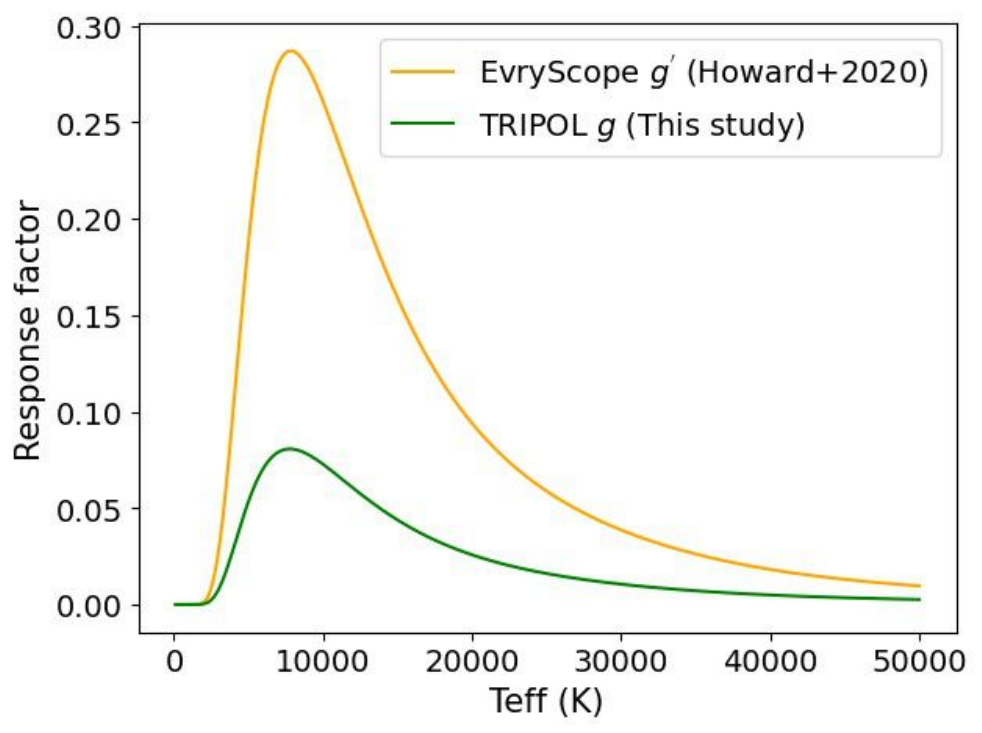}
    \caption{{TRIPOL $g$~band (green) response factor as the function of blackbody temperature. 
    The Evryscope $g^{'}$~band response factor function (orange) from \cite{2020ApJ...902..115H} is displayed for comparison.}
    }
    \label{fig:tripol_response_factor_function}
\end{figure}

\begin{figure}
    \centering
    \figurenum{7}
    \epsscale{1}
    \plotone{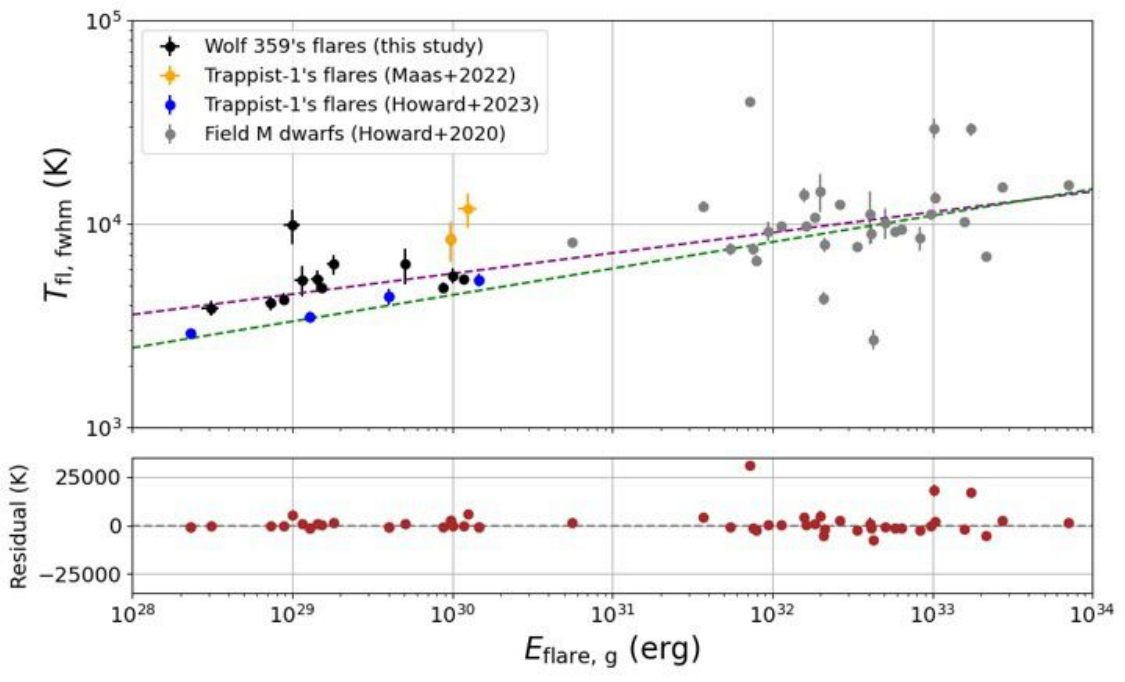}
    \caption{ 
    {Relationship between TRIPOL~$g$~band flare energies and FWHM temperatures for the M~dwarfs' flares. 
    {Black dots represent flares observed on Wolf~359 in this study.  
    Gray dots are flares from field M~dwarfs reported by \citet{2020ApJ...902..115H}.  
    TRAPPIST-1 flares are shown as yellow and blue dots, corresponding to data from \citet{Maas+2022} and \citet{2023ApJ...959...64H}, respectively.  
    The green dashed lines show the power-law correlations derived from M~dwarf superflares reported by \citet{2020ApJ...902..115H}. 
    The purple dashed lines represent the power-law correlations derived from low-energy Wolf~359's flares reported in this study and Trappist-1's flares reported in \citet{Maas+2022} and \citet{2023ApJ...959...64H} (see Equation~\ref{eq:flare_energy_temperature} and the coefficients in Table~\ref{tab:flare_temperature-vs-energy_powerlaw}).
    The subpanels below each panel show the residuals diagrams relative to the solar-class flare power-laws, visually demonstrating the goodness of fit for flares across all energy regimes.}}}
    \label{fig:flare_temperature_vs_energy}
\end{figure}


\begin{figure}
    \centering
    \figurenum{8}
    \epsscale{0.7}
    \plotone{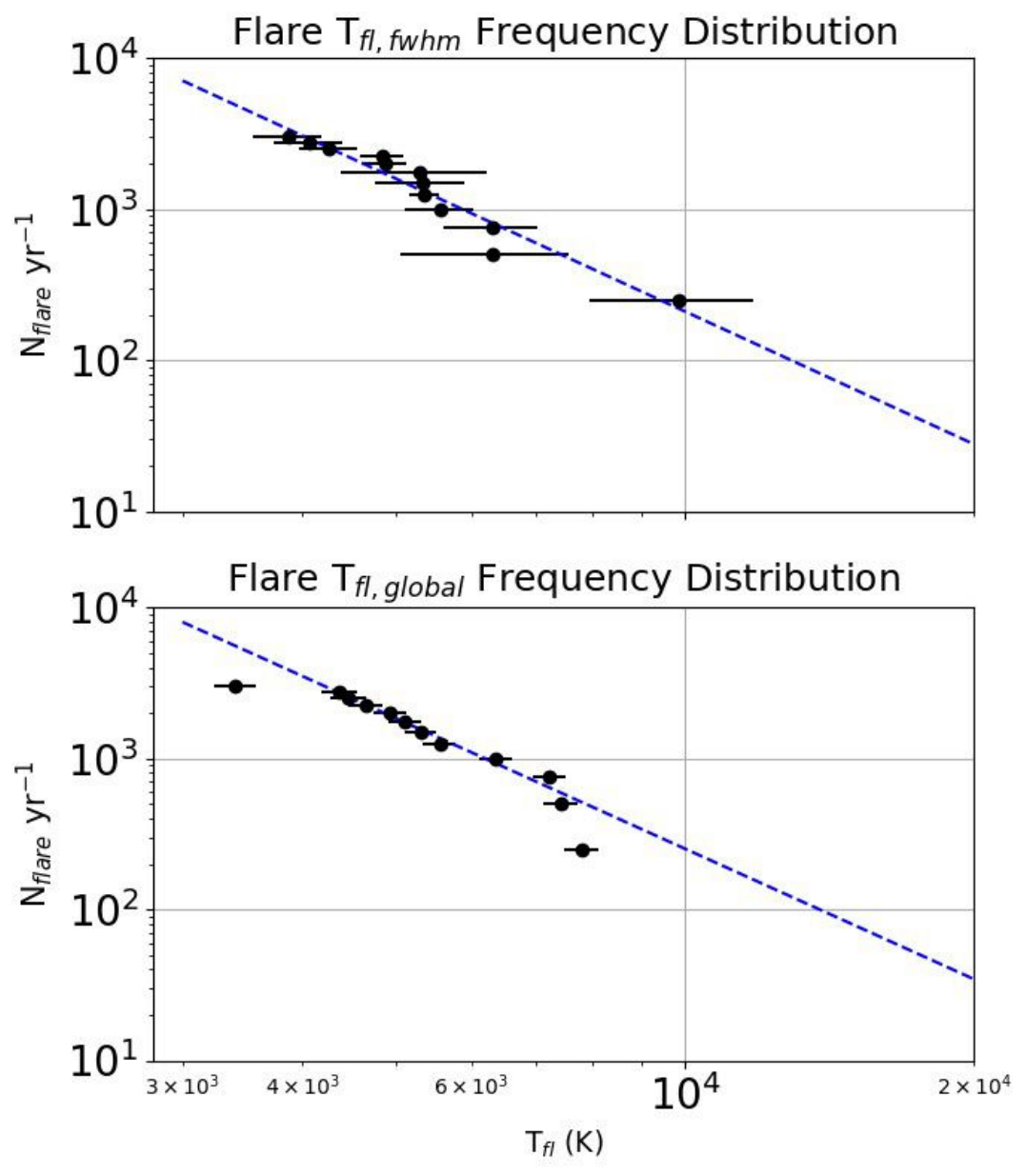}
    \caption{
    Cumulative frequency distributions of flare temperatures for Wolf~359.  
    \textbf{Top:} Flare FWHM temperature distribution ($T_{\mathrm{fl},\mathrm{fwhm}}$).  
    \textbf{Bottom:} Global flare temperature distribution ($T_{\mathrm{fl},\mathrm{glob}}$).  
    In both panels, the cumulative number of flares per year is plotted as a function of temperature, with the blue dashed lines showing the best-fit power-law relations.  
    The fitted slopes are {$-2.92 \pm 0.31$ for $T_{\mathrm{fl},\mathrm{fwhm}}$ and $-2.87 \pm 0.21$ for $T_{\mathrm{fl},\mathrm{glob}}$}.
    {We note that these empirically derived slopes of the flare temperature frequency distributions may be biased and shallower than what they should be due to the limited sample size. A more complete and statistically robust flare dataset will be crucial for more accurately constraining its underlying power-law behavior.}
    }
    \label{fig:flare_temperature_frequency_distribution}
\end{figure}

\begin{figure}
    \centering
    \figurenum{9}
    \epsscale{0.5}
    \plotone{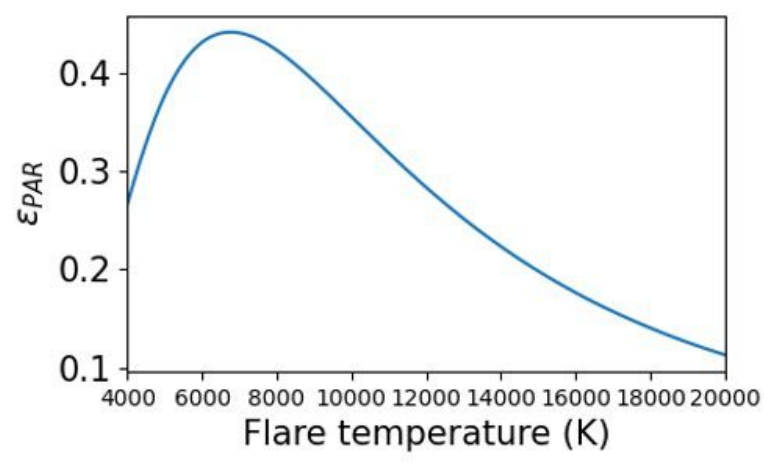}
    \caption{
    The fraction of bolometric flare energy emitted within the PAR wavelength range as function of flare temperature, $\epsilon_{\mathrm{PAR}}(T_{\mathrm{fl}})$, we employ in Eq.~\ref{eq:Lingam}. 
    }
    \label{fig:epsilon_PAR}
\end{figure}

\clearpage

\clearpage

\startlongtable
\begin{deluxetable}{lcccccccccc}
\tablecaption{Stellar Properties of Wolf~359}
\tablecolumns{11}
\tablewidth{0pt}
\tablehead{
\colhead{Name} & 
\colhead{SpT} & 
\colhead{$d$} & 
\colhead{$R_*$} & 
\colhead{$T_{\rm eff}$} & 
\colhead{$P_{\rm rot}$} & 
\colhead{Age} & 
\colhead{$g$} & 
\colhead{$r$} & 
\colhead{$i$} & 
\colhead{Flare Rate} \\
\multicolumn{1}{c}{} & 
\multicolumn{1}{c}{} & 
\multicolumn{1}{c}{(pc)} & 
\multicolumn{1}{c}{($R_\odot$)} & 
\multicolumn{1}{c}{(K)} & 
\multicolumn{1}{c}{(days)} & 
\multicolumn{1}{c}{(Gyr)} & 
\multicolumn{1}{c}{(mag)} & 
\multicolumn{1}{c}{(mag)} & 
\multicolumn{1}{c}{(mag)} & 
\multicolumn{1}{c}{(day$^{-1}$)}
}
\startdata
Wolf~359 & M5.5--M6 & 2.42 & 0.16 & 2700--2900 & $2.72 \pm 0.04$ & 0.1--1.5 & 14.265 & 12.762 & 10.320 & $\sim$~11 \\
\enddata
\tablecomments{
\textbf{Reference of every parameter:}
Spectral type (SpT): \citet{1995AJ....110.1838R,1991ApJS...77..417K}. 
Distance ($d$): \citet{2016AJ....152...24W}. 
Stellar radius ($R_*$): \cite{stellar_radius}.
Effective temperature ($T_{\rm eff}$): \citet{effective_temperature_1, effective_temperature_2}.
Rotation period ($P_{\rm rot}$): \citet{2018RNAAS...2....1G}.
Stellar age: \citet{2023AJ....166..260B}.
Magnitude in the $g$, $r$, and $i$: \citet{2012yCat.1322....0Z}.
Flare occurrence rate: \citet{2021AJ....162...11L}.
}
\label{tab:wolf359_summary}
\end{deluxetable}

\begin{deluxetable*}{c c c c c c c}
\label{tab:obs_summary}
\tabletypesize{\normalsize}
\tablecaption{\\Summary of the photometric and spectroscopic observations of Wolf~359.}
\tablecolumns{7}
\tablewidth{0pt}
\tablehead{
\colhead{Date (UTC)} & 
\colhead{Total Obs. Time (hr)} &
\colhead{Telescope} & 
\colhead{Instrument} & 
\colhead{Exposure} & 
\colhead{Cadence} & 
\colhead{Observation Type}
}
\startdata
2023 Feb 17 (13:01:38–21:39:43) & 8.63 & Lulin 1-m   & TRIPOL      & 30 sec      & $\sim$35 sec  & Photometry ($g$, $r$, $i$) \\
                                &  & Lulin 41-cm & Photometer  & 300s ($u$), 5s ($z$) & $\sim$364 sec & Photometry ($u$, $z$) \\
\hline
2023 Feb 18 (12:27:16–21:42:17) & 9.25 & Lulin 1-m   & TRIPOL      & 30 sec      & $\sim$35 sec  & Photometry ($g$, $r$, $i$) \\
                                &  & Lulin 41-cm & Photometer  & 5 sec       & $\sim$20 sec  & Photometry ($z$ only) \\
\hline
2023 Feb 20 (12:18:57–21:11:40) & 8.88 & Lulin 1-m   & TRIPOL      & 20 sec      & $\sim$25 sec  & Photometry ($g$, $r$, $i$) \\
                                &  & Lulin 41-cm & Photometer  & 5 sec       & $\sim$20 sec  & Photometry ($z$ only) \\
\hline
2023 Feb 21 (12:13:37–21:32:41) & 9.19 & Lulin 1-m   & TRIPOL      & 20 sec      & $\sim$25 sec  & Photometry ($g$, $r$, $i$) \\
                                &  & Lulin 41-cm & Photometer  & 5 sec       & $\sim$20 sec  & Photometry ($z$ only) \\
\hline
2023 Feb 22 (12:08:33–21:23:19) & 9.25 & Lulin 1-m   & TRIPOL      & 20 sec      & $\sim$25 sec  & Photometry ($g$, $r$, $i$) \\
                                &  & Lulin 41-cm & Photometer  & 5 sec       & $\sim$20 sec  & Photometry ($z$ only) \\
\hline
2021 Apr 20 (12:02:08–15:48:34) & 3.77 & Lulin 1-m   & LISA spectrograph & 600 sec & $\sim$605 sec       & Spectroscopy (4000--8400~\AA) \\
                                &  & Lulin 41-cm & Photometer       & 180 sec & $\sim$190 sec & Photometry ($u$ only) \\
\enddata
\tablecomments{
{The total cadence includes both the exposure time and the instrument readout time. For the observations on 2023 February 17, the cadence also accounts for the filter wheel switching time.
}
}
\end{deluxetable*}



\movetabledown=6.5cm
\begin{rotatetable}
\begin{deluxetable}{c c c c c c c c c c c c}
\tabletypesize{\normalsize}

\tablecaption{\\Properties of Wolf~359's flares detected in this study}
\tablecolumns{12}
\tablewidth{0pt}
\tablehead{
\colhead{Flare ID} & 
\colhead{Time of peak} & 
\colhead{$\tau_u$} & 
\colhead{$\tau_{g\&r}$} &
\colhead{$A_{f,u}$} & 
\colhead{$A_{f,g}$} & 
\colhead{$A_{f,r}$} & 
\colhead{$A_{f,i}$} & 
\colhead{$E_{f,u}$} & 
\colhead{$E_{f,g}$} & 
\colhead{$E_{f,r}$} & 
\colhead{$E_{f,i}$} \\
\colhead{} & 
\colhead{{MJD}} & 
\colhead{(min)} & 
\colhead{(min)} & 
\colhead{} & 
\colhead{} & 
\colhead{} & 
\colhead{} & 
\colhead{($10^{28}$~erg)} & 
\colhead{($10^{28}$~erg)} & 
\colhead{($10^{28}$~erg)} & 
\colhead{($10^{28}$~erg)} \\
\colhead{(1} & 
\colhead{{2}} & 
\colhead{(3)} & 
\colhead{(4)} & 
\colhead{(5)} & 
\colhead{(6)} & 
\colhead{(7)} & 
\colhead{(8)} & 
\colhead{(9)} & 
\colhead{(10)} & 
\colhead{(11)} & 
\colhead{(12)} 
}
\startdata
F1-0217 & 59992.577274 & 12.13 & 3.50 & 2.432 $\pm$ 0.209 & 0.267 $\pm$ 0.022 & 0.073 $\pm$ 0.008 & 0.005 $\pm$ 0.005 & 22.10 $\pm$ 3.44 & 18.00 $\pm$ 1.45 & 12.20 $\pm$ 1.15 & \nodata \\
F2-0217 & 59992.620376 & 24.27 & 2.92 & 1.506 $\pm$ 0.152 & 0.169 $\pm$ 0.020 & 0.038 $\pm$ 0.007 & \nodata & 32.10 $\pm$ 3.81 & 10.00 $\pm$ 1.16 & 7.04 $\pm$ 0.95 & \nodata \\
F3-0217 & 59992.786453 & 36.42 & 3.52 & 2.388 $\pm$ 0.178 & 0.229 $\pm$ 0.019 & 0.086 $\pm$ 0.007 & \nodata & 54.40 $\pm$ 5.39 & 11.50 $\pm$ 1.20 & 12.20 $\pm$ 0.98 & \nodata \\
F4-0218 & 59993.851221 & \nodata & 24.85 & \nodata & 0.248 $\pm$ 0.026 & 0.086 $\pm$ 0.009 & \nodata & \nodata & 117.00 $\pm$ 4.19 & 152.00 $\pm$ 3.31 & \nodata \\
F5-0220 & 59995.542147 & \nodata & 2.92 & \nodata & 0.266 $\pm$ 0.021 & 0.095 $\pm$ 0.007 & 0.013 $\pm$ 0.005 & \nodata & 14.30 $\pm$ 1.28 & 13.70 $\pm$ 0.91 & 19.40 $\pm$ 6.37 \\
F6-0220 & 59995.793964 & \nodata & 12.30 & \nodata & 0.199 $\pm$ 0.032 & 0.078 $\pm$ 0.011 & \nodata & \nodata & 50.90 $\pm$ 3.19 & 41.64 $\pm$ 2.57 & \nodata \\
F7-0221 & 59996.606464 & \nodata & 24.53 & \nodata & 0.196 $\pm$ 0.013 & 0.078 $\pm$ 0.003 & \nodata & \nodata & 87.74 $\pm$ 2.21 & 98.00 $\pm$ 1.18 & \nodata \\
F8-0221 & 59996.621233 & \nodata & 2.52 & \nodata & 0.083 $\pm$ 0.011 & 0.050 $\pm$ 0.003 & \nodata & \nodata & 7.36 $\pm$ 0.55 & 9.85 $\pm$ 0.30 & \nodata \\
F9-0221 & 59996.675573 & \nodata & 2.50 & \nodata & 0.083 $\pm$ 0.008 & 0.035 $\pm$ 0.002 & \nodata & \nodata & 3.09 $\pm$ 0.40 & 6.54 $\pm$ 0.24 & \nodata \\
F10-0221 & 59996.727193 & \nodata & 3.70 & \nodata & 0.087 $\pm$ 0.008 & 0.045 $\pm$ 0.002 & 0.005 $\pm$ 0.001 & \nodata & 8.82 $\pm$ 0.49 & 10.73 $\pm$ 0.30 & 26.57 $\pm$ 0.62 \\
F11-0221 & 59996.884566 & \nodata & 4.10 & \nodata & 0.179 $\pm$ 0.012 & 0.068 $\pm$ 0.003 & \nodata & \nodata & 15.33 $\pm$ 0.86 & 15.51 $\pm$ 0.44 & \nodata \\
F12-0222 & 59997.541522 & \nodata & 11.77 & \nodata & 0.826 $\pm$ 0.063 & 0.300 $\pm$ 0.015 & 0.045 $\pm$ 0.012 & \nodata & 100.11 $\pm$ 4.23 & 72.27 $\pm$ 2.67 & 14.64 $\pm$ 8.16 \\
\enddata
\tablecomments{
{The columns are: 
(1) Assigned Flare ID; 
(2) Time of flare peak in MJD; 
(3) Flare duration in the $u$ band ($\tau_u$);
(4) Flare duration in the $g$ and $r$ bands ($\tau_{g\&r}$);
(5–8) Flare amplitudes ($A_{f,x=band}$) in the $u$, $g$, $r$, and $i$ bands, respectively;
(9–12) Flare energies ($E_{f,x=band}$) in units of $10^{28}$ erg for the $u$, $g$, $r$, and $i$ bands, respectively.
}}
\label{tab:flare_properties}
\end{deluxetable}
\end{rotatetable}

\begin{deluxetable}{c c c c c}
\label{tab:flare_temperature_comparison}
\tabletypesize{\normalsize}
\tablecaption{\\Comparison of maximum flare temperatures derived from SED fitting and flux ratios for selected flares.}
\tablecolumns{5}
\tablewidth{0pt}
\tablehead{
\colhead{Flare ID} & 
\colhead{$T_{fl,\mathrm{SED}}$ (K)} & 
\colhead{$T_{fl,gr}$ (K)} & 
\colhead{$T_{fl,ri}$ (K)} & 
\colhead{$T_{fl,gi}$ (K)}
}
\startdata
F5-0220 & $5700 \pm 600$ & $5900 \pm 700$ & $4000 \pm 600$ & $5000 \pm 300$ \\
F10-0221 & $4500 \pm 400$ & $4500 \pm 500$ & $4800 \pm 600$ & $4600 \pm 300$ \\
F12-0222 & $5500 \pm 500$ & $5800 \pm 700$ & $3700 \pm 500$ & $4700 \pm 300$ \\
\enddata
\end{deluxetable}

\begin{deluxetable}{c c c c c}
\label{tab:flare_temp_xbb}
\tabletypesize{\normalsize}
\tablecaption{\\Blackbody FWHM and global flare temperatures and filling factors determined by using $g$/$r$ color ratio.}
\tablecolumns{5}
\tablewidth{0pt}
\tablehead{
\colhead{Flare ID} & 
\colhead{FWHM (min)} &
\colhead{$T_{fl,\mathrm{fwhm}}$ (K)} &
\colhead{$T_{fl,\mathrm{glob}}$ (K)} & 
\colhead{$X_{BB}$ (ppm)}
}
\startdata
F1-0217 & 0.933 & 6300 $\pm$ 700 & 7800 $\pm$ 300 & 33.68 $\pm$ 2.78 \\
F2-0217 & 0.839 & 9900 $\pm$ 1900 & 7400 $\pm$ 300 & 8.35 $\pm$ 0.97 \\
F3-0217 & 0.696 & 5300 $\pm$ 900 & 5100 $\pm$ 200 & 130.83 $\pm$ 10.95 \\
F4-0218 & 0.497 & 5400 $\pm$ 200 & 4500 $\pm$ 200 & 95.57 $\pm$ 10.17 \\
F5-0220 & 0.855 & 5300 $\pm$ 600 & 5600 $\pm$ 200 & 115.89 $\pm$ 9.29 \\
F6-0220 & 1.921 & 6300 $\pm$ 1300 & 6300 $\pm$ 200 & 133.45 $\pm$ 21.49 \\
F7-0221 & 1.207 & 4900 $\pm$ 300 & 4900 $\pm$ 200 & 142.15 $\pm$ 9.41 \\
F8-0221 & 1.811 & 4100 $\pm$ 300 & 4400 $\pm$ 200 & 368.71 $\pm$ 47.43 \\
F9-0221 & 0.543 & 3900 $\pm$ 300 & 3400 $\pm$ 200 & 77.24 $\pm$ 7.10 \\
F10-0221 & 1.607 & 4300 $\pm$ 300 & 4700 $\pm$ 200 & 202.17 $\pm$ 17.64 \\
F11-0221 & 0.845 & 4900 $\pm$ 300 & 5300 $\pm$ 200 & 105.86 $\pm$ 7.30 \\
F12-0222 & 0.483 & 5600 $\pm$ 500 & 7200 $\pm$ 300 & 395.72 $\pm$ 30.24 \\
\hline
Average & 1.02 & 5500 $\pm$ 1600 & 5600 $\pm$ 1400 & 150.80 $\pm$ 119.31 \\
\enddata
\end{deluxetable}

\begin{deluxetable}{ccccl}
\label{tab:m_dwarfs_flares_in_temperature_energy_figure}
\tabletypesize{\footnotesize}
\tablecaption{Temperatures and energies (in TRIPOL $g$~band and in bolometric) of flares from various M dwarf sources those are shown in Figure~\ref{fig:flare_temperature_vs_energy}}
\tablecolumns{5}
\tablewidth{0pt}
\tablehead{
\colhead{$T_{fl,\mathrm{fwhm}}$ (K)} & 
\colhead{$\log_{10} E_{g} (erg)$} & 
\colhead{$\log_{10} E_{\mathrm{bol}}(erg)$} & 
\colhead{Sources}
}
\startdata
3900 $\pm$ 300 & 28.49 & 30.2 & Wolf 359 (this study) \\
4100 $\pm$ 300 & 28.87 & 30.49 & Wolf 359 (this study) \\
4300 $\pm$ 300 & 28.95 & 30.51 & Wolf 359 (this study) \\
4800 $\pm$ 300 & 29.19 & 30.6 & Wolf 359 (this study) \\
4900 $\pm$ 300 & 29.94 & 31.35 & Wolf 359 (this study) \\
5300 $\pm$ 900 & 29.06 & 30.39 & Wolf 359 (this study) \\
5300 $\pm$ 600 & 29.16 & 30.48 & Wolf 359 (this study) \\
5400 $\pm$ 200 & 30.07 & 31.39 & Wolf 359 (this study) \\
5600 $\pm$ 500 & 30.00 & 31.30 & Wolf 359 (this study) \\
6300 $\pm$ 700 & 29.26 & 30.49 & Wolf 359 (this study) \\
6300 $\pm$ 1300 & 29.71 & 30.94 & Wolf 359 (this study) \\
9900 $\pm$ 1900 & 29.00 & 30.22 & Wolf 359 (this study) \\
11800 $\pm$ 2300 & 30.10 & 31.48 & Trappist-1 \citep{Maas+2022} \\
8400 $\pm$ 1900 & 29.99 & 31.18 & Trappist-1 \citep{Maas+2022} \\
5300 $\pm$ 400 & 30.17 & 31.50 & Trappist-1 \citep{2023ApJ...959...64H} \\
4400 $\pm$ 400 & 29.60 & 31.12 & Trappist-1 \citep{2023ApJ...959...64H} \\
3500 $\pm$ 200 & 29.11 & 31.00 & Trappist-1 \citep{2023ApJ...959...64H} \\
2900 $\pm$ 100 & 28.37 & 30.69 & Trappist-1 \citep{2023ApJ...959...64H} \\
15500 $\pm$ 400 & 33.85 & 35.77 & Field M dwarfs \citep{2020ApJ...902..115H} \\
15100 $\pm$ 300 & 33.44 & 35.35 & Field M dwarfs \citep{2020ApJ...902..115H} \\
6900 $\pm$ 300 & 33.34 & 34.99 & Field M dwarfs \citep{2020ApJ...902..115H} \\
7700 $\pm$ 300 & 32.53 & 34.17 & Field M dwarfs \citep{2020ApJ...902..115H} \\
13400 $\pm$ 800 & 33.02 & 34.85 & Field M dwarfs \citep{2020ApJ...902..115H} \\
2700 $\pm$ 300 & 32.63 & 35.62 & Field M dwarfs \citep{2020ApJ...902..115H} \\
29200 $\pm$ 2200 & 33.24 & 35.74 & Field M dwarfs \citep{2020ApJ...902..115H} \\
9200 $\pm$ 300 & 32.76 & 34.43 & Field M dwarfs \citep{2020ApJ...902..115H} \\
29500 $\pm$ 3200 & 33.01 & 35.52 & Field M dwarfs \citep{2020ApJ...902..115H} \\
11200 $\pm$ 3300 & 32.61 & 34.34 & Field M dwarfs \citep{2020ApJ...902..115H} \\
9400 $\pm$ 300 & 32.80 & 34.47 & Field M dwarfs \citep{2020ApJ...902..115H} \\
8500 $\pm$ 1100 & 32.92 & 34.57 & Field M dwarfs \citep{2020ApJ...902..115H} \\
12400 $\pm$ 700 & 32.42 & 34.20 & Field M dwarfs \citep{2020ApJ...902..115H} \\
39700 $\pm$ 2000 & 31.86 & 34.70 & Field M dwarfs \citep{2020ApJ...902..115H} \\
10200 $\pm$ 400 & 33.20 & 34.89 & Field M dwarfs \citep{2020ApJ...902..115H} \\
11100 $\pm$ 400 & 32.99 & 34.72 & Field M dwarfs \citep{2020ApJ...902..115H} \\
10700 $\pm$ 300 & 32.26 & 33.97 & Field M dwarfs \citep{2020ApJ...902..115H} \\
10100 $\pm$ 1700 & 32.70 & 34.39 & Field M dwarfs \citep{2020ApJ...902..115H} \\
13900 $\pm$ 1100 & 32.20 & 34.05 & Field M dwarfs \citep{2020ApJ...902..115H} \\
8900 $\pm$ 900 & 32.61 & 34.27 & Field M dwarfs \citep{2020ApJ...902..115H} \\
9800 $\pm$ 500 & 32.21 & 33.89 & Field M dwarfs \citep{2020ApJ...902..115H} \\
14500 $\pm$ 3000 & 32.30 & 34.18 & Field M dwarfs \citep{2020ApJ...902..115H} \\
7900 $\pm$ 600 & 32.33 & 33.97 & Field M dwarfs \citep{2020ApJ...902..115H} \\
6600 $\pm$ 400 & 31.90 & 33.56 & Field M dwarfs \citep{2020ApJ...902..115H} \\
9200 $\pm$ 1000 & 31.97 & 33.64 & Field M dwarfs \citep{2020ApJ...902..115H} \\
7500 $\pm$ 500 & 31.74 & 33.38 & Field M dwarfs \citep{2020ApJ...902..115H} \\
9700 $\pm$ 600 & 32.05 & 33.73 & Field M dwarfs \citep{2020ApJ...902..115H} \\
7500 $\pm$ 500 & 31.88 & 33.52 & Field M dwarfs \citep{2020ApJ...902..115H} \\
4300 $\pm$ 300 & 32.32 & 34.32 & Field M dwarfs \citep{2020ApJ...902..115H} \\
12200 $\pm$ 800 & 31.57 & 33.34 & Field M dwarfs \citep{2020ApJ...902..115H} \\
8100 $\pm$ 400 & 30.75 & 32.39 & Field M dwarfs \citep{2020ApJ...902..115H} \\
\enddata
\tablecomments{
{Please read Section~\ref{subsect:flare_temperature_and_energy} for the detailed information about the flare energy conversion between different bandpasses.
}}
\end{deluxetable}


\begin{deluxetable}{l c c c}
\label{tab:flare_temperature-vs-energy_powerlaw}
\tabletypesize{\normalsize}
\tablecaption{\\Power-law coefficients ($\alpha$ and $\beta$) of flare FWHM temperatures (T$_{\mathrm{fl},\mathrm{fwhm}}$) versus flare energies ($E_{f,\lambda}$) in the TRIPOL~$g$ band for the Eq.~\ref{eq:flare_energy_temperature}, derived separately for superflares and solar-class flares.}
\tablecolumns{4}
\tablewidth{0pt}
\tablehead{
\colhead{Flare Class} &
\colhead{$E_{f,\lambda}$} &
\colhead{$\alpha$} &
\colhead{$\beta$}
}
\startdata
\textbf{Superflares} &      $\lambda = \mathrm{EvryScope\ }g'$ & $0.128^{a}$ & $-0.193^{a}$ \\
                          & $\lambda = \mathrm{TRIPOL\ }g$ & $0.130 \pm 0.06$ & $-2.501 \pm 1.051$ \\
\hline
\textbf{Solar-class flares} & $\lambda = \mathrm{TRIPOL\ }g$ & $0.103 \pm 0.045$ & $0.655 \pm 0.512$ \\
\enddata
\tablecomments{
{$^{a}$ Reported by \citet{2020ApJ...902..115H}.
The superflare sample is taken from field M~dwarfs reported by \citet{2020ApJ...902..115H}.
The solar-class flare sample combines the Wolf~359 flares observed in this study with 
the Trappist-1 flares reported by \citet{Maas+2022} and \citet{2023ApJ...959...64H}. 
The visual illustrations are displayed in Figure~\ref{fig:flare_temperature_vs_energy}.
}}
\end{deluxetable}


%
\clearpage
\vspace{5mm}
\facilities{Lulin 1-m telescope (LOT), Lulin 41-cm telescope (SLT)}


\software{\texttt{Astropy} \citep{2013A&A...558A..33A,2018AJ....156..123A}, \texttt{NumPy} \citep{harris2020array}, \texttt{Scipy} \citep{2020NatMe..17..261V}, \texttt{Matplotlib} \citep{2007CSE.....9...90H}, \texttt{Pandas} \citep{mckinney2010data},
\texttt{pymc3} \citep{2016ascl.soft10016S}
}





\bibliography{sample631}{}
\bibliographystyle{aasjournal}



\end{document}